\documentclass[journal]{IEEEtran}
\usepackage[utf8]{inputenc}
\usepackage[english]{babel}
\usepackage{multicol}               
\usepackage[a4paper, margin = 0.75in]{geometry}      

\usepackage{mathtools} 
\usepackage{amsmath}
\usepackage{graphicx} 
\usepackage{float}
\usepackage{fancyhdr} 

\usepackage{comment}
\usepackage[table]{xcolor}
\usepackage{multirow}

\usepackage{caption}
\usepackage{subcaption}

\usepackage[ruled,vlined,linesnumbered]{algorithm2e}

\usepackage{hyperref}
\usepackage{csquotes}
\usepackage{cleveref}
\usepackage{makecell}
\usepackage{multicol}

\renewcommand\IEEEkeywordsname{Keywords}

\title{Reliability Assessment for Distribution Systems with Embedded Microgrids} 
\author{
\IEEEauthorblockN{Stine Fleischer Myhre\IEEEauthorrefmark{1}, Olav Bjarte Fosso\IEEEauthorrefmark{1}, Oddbjørn Gjerde\IEEEauthorrefmark{2}, Poul Einar Heegaard\IEEEauthorrefmark{3}} \\
\IEEEauthorblockA{\IEEEauthorrefmark{1}Department of Electric Power Engineering, NTNU\\
\IEEEauthorrefmark{2}SINTEF Energy Research\\
\IEEEauthorrefmark{3}Dept. of Information Security and Communication Technology, NTNU  \\
Trondheim, Norway\\
\{stine.f.myhre, olav.fosso, poul.heegaard\}@ntnu.no; oddbjorn.gjerde@sintef.no}
}
\date{November 2021}

\begin{document}
\maketitle

\begin{abstract}
\label{abstract}
    The electrical distribution system is moving toward a more decentralized, complex, and dynamic system. The system is experiencing a higher penetration of distributed energy resources, flexible resources, and active end-users, leading to an active distribution system. If these components are used actively, they can have a positive effect on the distribution system's reliability. 
    This paper aims to investigate how a microgrid with renewable energy sources and energy storage might influence the reliability of electricity supply in a radially operated distribution system. The reliability is investigated from both the distribution system and the microgrid perspective with the application of different scenarios. The study includes an extensive sensitivity analysis. In this study, we use our developed reliability assessment tool for distribution networks. The tool is an open available software where the example network and datasets are embedded. The reliability assessment tool is based on Monte Carlo simulations where load flow calculations are included to capture the behavior of the system and system components. In this paper, the reliability tool is extended to include microgrids and microgrid controller opportunities that handle interaction strategies with the distribution network. The model is demonstrated on the IEEE 33-bus network. The result is confirmed through statistical testing showing the statistical significance of providing support from the microgrid on the distribution system's reliability. 

\end{abstract}


\section{Introduction}
\label{Introduction}


The electrical power system is constantly developing, and in the upcoming years, a lot of changes will take place \cite{national2016power, IEA2019}. The power system is moving from being hierarchical with the one-way transfer of energy, to a more decentralized system where the system boundaries will be more invisible. The biggest changes can be seen in the MV distribution system, which will be more decentralized, complex, and dynamic with higher penetration of distributed energy resources and active end-users \cite{ipakchi2009, Prettico2019}. This will transform the distribution system into an active system with bidirectional power flows that will change the system's behavior.


From a reliability perspective, this creates new possible solutions. With more flexible resources in the system and integration of distributed generation (DG), there is a potential to increase the system's reliability. 
Microgrids are an increasingly studied concept related to the improvement of system reliability \cite{knowledge2019microgrid}. The main attribute of microgrids is the possibility to operate in two different modes: 1) \textit{Grid-connected mode} and 2) \textit{Islanded mode}. Thus, the microgrid can operate in an islanded mode during outages in the main system. If the microgrid and the distribution system work together, there is a potential for improved reliability related to unintentional outages in the system. By islanding parts of the distribution system with the microgrid, the microgrid could be the provider of these distribution system load points.

For microgrid support to work, an interaction between the distribution system operator (DSO) and the microgrid owner must be in place. The type of interaction is dependent on who owns the microgrid. In \cite{Thema, EUWinter}, and \cite{ton2012us}, this theme is discussed. In addition, regulatory frameworks related to the responsibility of islanded load points need to be established. The regulatory interaction frameworks will not be considered in this paper but are nevertheless important aspects to consider for the application of such support.




\subsection{Related work}

Some research has already been done to investigate the potential reliability improvement with technologies such as DG and flexible resources integrated into distribution systems. In \cite{sperstad2020impact}, an overview of how flexible resources may impact the security of electricity supply (SoS) is provided. The study also includes a review of methods and indicators to quantify the impact. A review of power system flexibility and flexibility concerning system security is provided in \cite{mohandes2019review}. 
A survey of reliability assessment techniques for modern distribution systems is assessed in \cite{escalera2018survey}. Here, multiple modeling techniques used for assessing distribution system reliability are investigated concerning active components in the system. The availability and impact of active components, such as DG \cite{borges2012overview, celli2013reliability}, energy storage units \cite{mohamad2018development}, and systems with a combination of these \cite{kwasinski2012availability, nosratabadi2017comprehensive}, have been investigated. The literature illustrates that DG and flexible resources might have a positive impact on the reliability of distribution systems if administrated wisely.  


 Some studies have investigated the possibility of islanding parts of the distribution system during failure events. In \cite{costa2009assessing} and \cite{conti2011generalized}, an analytical approach is adopted to evaluate the contribution from the microgrid to the distribution system reliability. The contribution of the microgrid is determined during possible operating conditions of the microgrid and the DG inside the microgrid and is evaluated based on the ratio between demand and supply. A simulation-based approach with Monte Carlo Simulations (MCS) has been used in \cite{conti2014monte} for intentional islanding of the distribution system. However, the paper does not consider the active participation of flexible resources and DG to restore supply. Furthermore, a sequential Monte Carlo Simulation (SMCS) method is used in \cite{de2017reliability} to evaluate the impact of local and mobile generation units. Here, parts of the system will be islanded, with some units operating as microgrids. In \cite{syrri2015contribution}, the reliability of a distribution system with support from a microgrid including combined heat and power supply system is assessed. However, the microgrid does not include renewable energy resources (RES). In addition, some research related to the reliability of multi-microgrids has been assessed in \cite{farzin2017role}, \cite{barani2018optimal}, and \cite{nikmehr2016reliability}. 
These papers do not consider the interaction between the microgrid and the distribution system, nor is the reliability of the microgrid investigated to any significant extent. 



\subsection{Contributions}

This paper investigates how a microgrid including RES may improve the reliability of a distribution system. We aim to map the reliability impact from the perspective of both the microgrid and the distribution network by varying the microgrid placement, operation mode, and capacity. Through the analysis, we aim to identify the optimal conditions for the microgrid to minimize the impact on the microgrid while maximizing the contributions to the distribution system. In addition, we provide a general methodology for evaluating how microgrids perform from a reliability perspective through the use of our developed reliability assessment tool, RELSAD.


We have created a reliability assessment tool, RELSAD (RELiability tool for Smart and Active Distribution networks), for assessing the reliability of modern distribution systems \cite{Myhre2022}. RELSAD is an open-source software with an associated documentation page \cite{RELSAD_documentation}. The documentation page records the example network with the datasets. 
In this paper, RELSAD is extended to include microgrids and the interaction between distribution systems and microgrids. Additionally, a microgrid controller with some different control modes is proposed. The presentation of the method focuses on the reliability analysis of a distribution system with the integration of a microgrid. The contributions of this paper are:

\begin{itemize}
    \item A methodology to assess the reliability of a distribution network with the active participation of a microgrid. Additionally, the reliability impact the microgrid experiences during the different operating modes of the microgrid is investigated.
    \item Demonstration of the method on a realistic case study where the reliability of both the distribution system and the microgrid is evaluated with different scenarios for microgrid support. The case study is conducted on the IEEE 33-bus network where distribution system data is gathered systematically through procedures described in Sec. \ref{ProcedureTestNetwork}. The reliability data of the distribution system is based on statistics from the Norwegian distribution systems. 
    \item A comprehensive sensitivity analysis of how microgrid battery capacity, line repair time, and line failure rate affects the Energy Not Supplied (ENS) of both the distribution system and the microgrid.
    \item An investigation of how the microgrid location affects the ENS of the distribution system.
\end{itemize}

\subsection{Paper structure}
The rest of this paper is organized as follows: 
\cref{SecurityofSupply} introduces the concept of failure handling in modern distribution systems with microgrids. In \cref{Reliability}, the method for calculating reliability in a distribution system with a microgrid is introduced and explained. The case study conducted on the IEEE 33-bus network is discussed in \cref{CaseStudy}. The case study results are illustrated and discussed in \cref{Results}. The sensitivity analysis and the investigation of the microgrid location is also presented in \cref{Results}. The paper is concluded, and future work is discussed, in \cref{Conclusion}.

\section{The reliability evaluation of the future distribution systems with microgrids}
\label{SecurityofSupply}

\begin{figure}[h]
    \centering
    \includegraphics[width=0.5\textwidth]{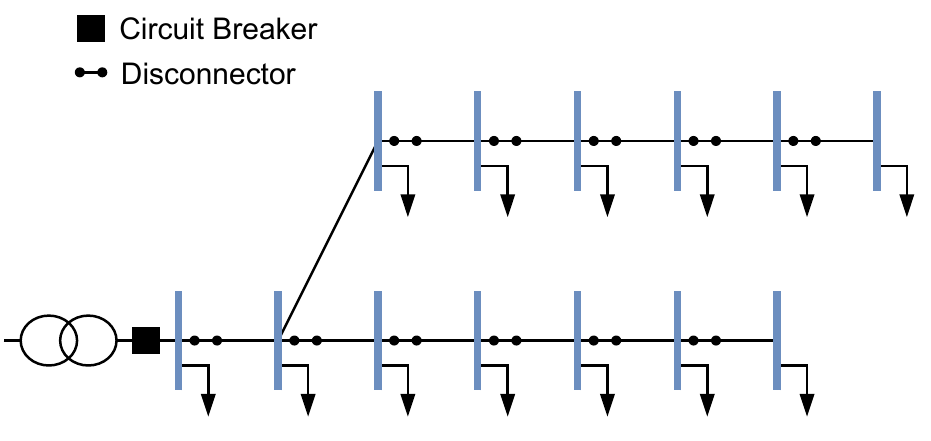}
  \caption{Traditional radially operated distribution system.}
  \label{fig:TradDist}
\end{figure}

\begin{figure}[h]
    \centering
    \includegraphics[width=0.5\textwidth]{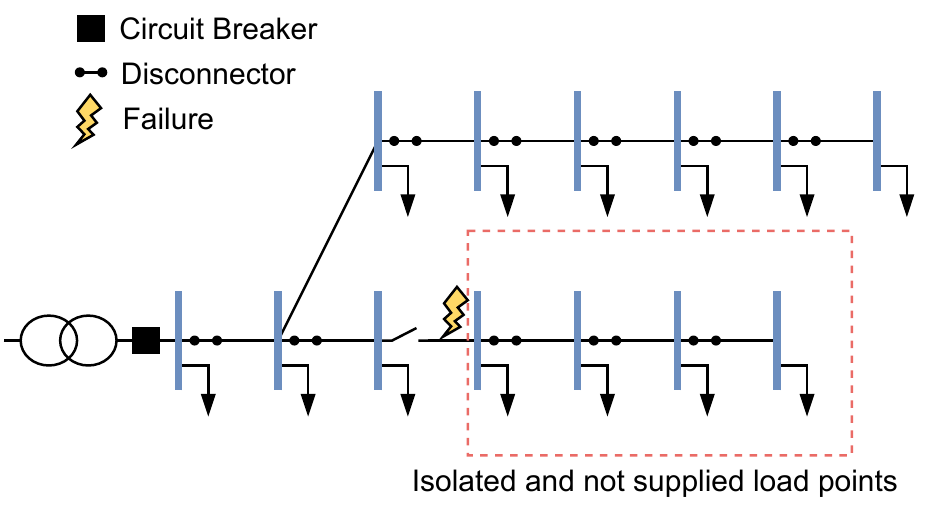}
  \caption{Traditional radially operated distribution system with a fault.}
  \label{fig:TradDistFailure}
\end{figure}

In a traditional radially operated distribution system with the one-way transfer of power, as seen in Fig. \ref{fig:TradDist}, a failure of, for example, a line in the network, will result in all the downstream load points being isolated from the main power source unless there is an alternative supply route. In Fig. \ref{fig:TradDistFailure}, a fault has occurred on a line in the distribution system and has been isolated. This results in the load points, highlighted in the red dotted box, downstream of the failed line, being isolated from the rest of the distribution system. Since there is no alternative supply route or any generation unit in this area, these load points will not be supplied until the failure is repaired and the line reconnected.   

However, if there is a flexible resource such as a microgrid present in the network, the situation might be changed. The aim is to use the flexibility in the system to restore or to support the network during periods with, e.g., faults in the network and capacity problems. A microgrid can support the system in two different modes: 1) \textit{isolated operation} and 2) \textit{grid-connected operation or supportive operation}. 

\subsubsection{Isolated operation}

\begin{figure}[h]
    \centering
    \includegraphics[width=0.5\textwidth]{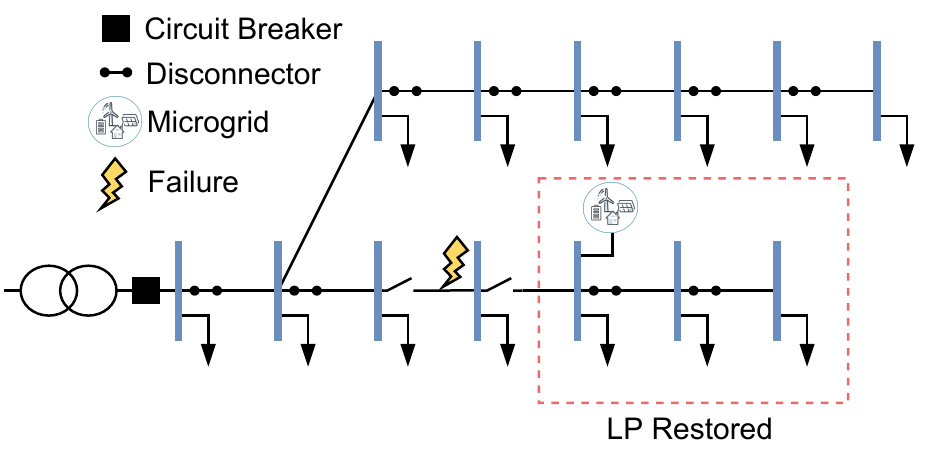}
  \caption{Supply restored for the microgrid in island mode with isolated part of the distribution syste.}
  \label{fig:MicrogridIsolated}
\end{figure}

If a failure occurs on, for example, a line in the distribution system as seen in Fig. \ref{fig:MicrogridIsolated}, the downstream load points will be isolated from the overlaying network and traditionally these load points will not be supplied. However, if a microgrid or another flexible resource is present in this isolated part of the network, at least some parts of the supply can be restored. The microgrid will then operate in an island mode during the outage time of the faulty component. When the fault is repaired, the isolated part can be disconnected from the overlying grid again. 

In this operation, the microgrid must be able to withhold the balance in the system along with the frequency and voltage levels in the isolated part of the network. This means that the microgrid is in charge of all the security of supply aspects during the period of isolated operation. 

\subsubsection{Supportive operation}

\begin{figure}[h]
    \centering
    \includegraphics[width=0.5\textwidth]{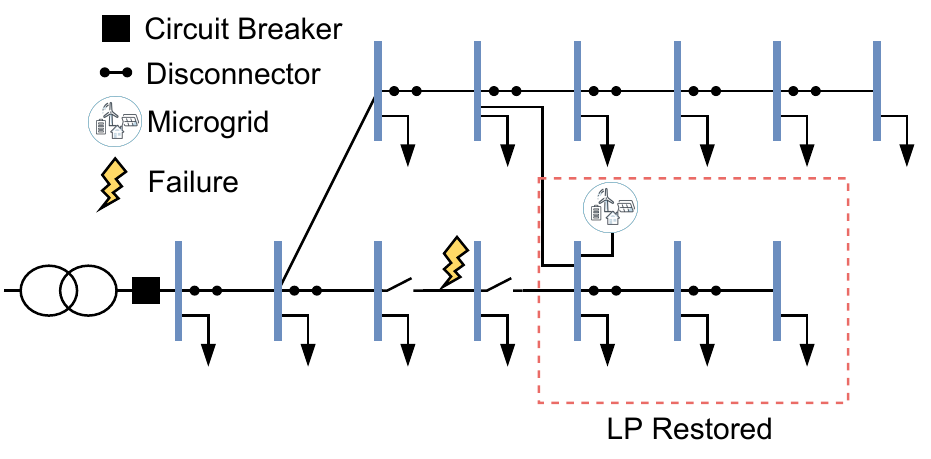}
  \caption{Supply restored with alternative supply chain and microgrid support.}
  \label{fig:MicrogridConnected}
\end{figure}

Normally, in distribution systems, some alternative supply chains might exist that can be used in case of redirection of power or failure in the network. Then the line can be connected and help supply loads in the system in areas that would have been isolated during the faulty period. However, some of these lines have low capacity limits and might end up being congested, or they might end up creating bottlenecks in the network. Local production in the distribution network might alleviate congestion and bottlenecks. Fig. \ref{fig:MicrogridConnected} illustrates how a microgrid can support the system during a fault in the system. The load points could have been restored with only the connection of the alternative line, but the microgrid is now able to support the system with local production. 


\section{Methodology}
\label{Reliability}

In this section, we present the developed reliability methodology. First, the concept for reliability evaluation of modern distribution systems is presented. Second, we introduce our developed reliability assessment tool with the extension to include microgrids and the interaction between microgrids and distribution networks. Finally, the statistical analysis used for evaluating simulation results is presented.

\subsection{Reliability evaluation of modern distribution systems}


In general, there are two ways of analyzing the reliability in the distribution system, namely through analytical approaches or simulation, with MCS frequently used \cite{Billinton}. The analytical approaches use simplified models with averaged quantities to estimate the reliability indices. MCS, however, allow for a more detailed representation of the reliability indices by generating reliability indices through numerous independent instances including stochastic variations. The numerous instances result in a detailed distribution of reliability indices. Analytical models are more generic and are often, therefore, more applicable for different cases, whereas MCS models will offer a more detailed analysis of the system to be modeled. 

To calculate the reliability of future smart and flexible distribution systems, an appropriate method is needed. The traditional approaches will not be able to utilize the full potential of the active components in the distribution system, and a form of simulation approach can be advantageous. Since the active components interact with the distribution system based on multiple independent factors, such as failures and weather conditions, the simulation of their impact on the distribution system must cover all possible scenarios. A MCS approach, which makes use of random sampling of input, is well suited for these types of problems. 

In this regard, RELSAD---RELiability tool for Smart and Active Distribution networks was developed. RELSAD is a reliability assessment tool for modern distribution systems with an SMCS model. The software is developed as an open-source Python package and is published as a scientific tool \cite{Myhre2022}. Additionally, a documentation page for the software has been written and can be seen in \cite{RELSAD_documentation}. RELSAD was developed to give a foundation for the reliability assessment of modern distribution systems where the changed behavior and the increased complexity in the distribution network are considered. The tool also aims to facilitate reliability calculation and analysis for the DSO. Through RELSAD, the following features are provided:

\begin{itemize}
    \item Calculation of detailed results in the form of distributions and statistics on important reliability indices.
    \item Opportunity for investigation of network sensitivities involving different parameters such as repair time, failure rate, load and generation profiles, placement of sources, and component capacities and their impact on the system.
    \item Implementation and investigation of diverse and advanced network constellations spanning from small passive distribution systems to large active networks including DG, batteries, and microgrids where the networks are operated radially.
    \item The ability to investigate and analyze the reliability of Cyber-Physical Networks.
\end{itemize}

In this paper, RELSAD is extended to include microgrids and interaction strategies between the microgrid and the distribution network during outages in the system. It will describe the core functionality of RELSAD in addition to the contribution and implementation of microgrids and the microgrid controller with the fault handling procedure. 



\subsection{Reliability indices}
\label{sec:rel_indices}

There are different reliability indices used for measuring and quantifying the reliability in distribution systems. Many of these indices are implemented in RELSAD. The reliability indices can be divided into \textit{customer-oriented indices} and \textit{load- and production-oriented indices} \cite{Billinton}. 
The three basic reliability parameters are the fault frequency or the average failure rate, $\lambda_s$, the annual average outage time, $U_s$, and the average outage time, $r_s$. In a radial network, the three different reliability parameters can be calculated as in eq. \ref{eq:faultfreq}, eq. \ref{eq:avgoutagetime}, and eq. \ref{eq:outagetime}. Here, $\lambda_i$ and $r_i$ are the failure rate and outage time at load point $i$, respectively. These equations are used to assess the load-point reliability. 
However, these do not contain any information about the electrical consequence of a fault or the cost related to a fault. The rest of this section will describe the reliability indices used in this paper. 
 
\begin{equation}
    \label{eq:faultfreq}
    \lambda_{s} = \sum \lambda_{i}
\end{equation} 

\begin{equation}
    \label{eq:avgoutagetime}
    U_{s} = \sum \lambda_{i} r_{i}
\end{equation}

\begin{equation}
    \label{eq:outagetime}
    r_{s} = \frac{U_{s}}{\lambda_{s}} = \frac{\sum \lambda_i r_i}{\sum \lambda_i}
\end{equation}

\subsubsection*{\bf Load- and Production-oriented indices}
The load- and production-oriented indices aim to indicate the electrical consequence of faults in the system \cite{Billinton}. The total ENS in a system can be calculated as seen in eq. \ref{eq:ENS} 

\begin{equation}
    \label{eq:ENS}
    {\tt ENS}_{s} = U_{s}P_{s}
\end{equation}


The interruption cost for the system can be calculated as seen in eq. \ref{eq:cost} \cite{943073}. 
Here, $c_{i}$ is the specific interruption cost for each customer category at load point $i$. 

\begin{equation}
    \label{eq:cost}
    {\tt CENS}_{s} = \sum {\tt ENS}_{i}c_{i}
\end{equation}

\subsubsection*{\bf Customer-oriented indices}

The customer-oriented indices aim to indicate the reliability of the distribution system based on the interruption experienced by the customers \cite{Billinton}. In this paper, three important indices are investigated:

\begin{enumerate}
    \item System Average Interruption Frequency Index
\begin{equation}
    \label{eq:SAIFI}
    {\tt SAIFI} = \frac{\sum_{\forall i} \lambda_{i}N_{i}}{\sum N_{i}}
\end{equation}
where $N_{i}$ is the {\em total number of customers served}, and $\sum_{\forall i} \lambda_{i}N_{i}$ is the {\em total number of customer interruptions}. ${\tt SAIFI}$ is a measure of the frequency of interruptions the customers in the system expect to experience. Any interruption seen from the consumer is counted as a fault, regardless of origin. 
\newline
\item System Average Interruption Duration Index
\begin{equation}
    \label{eq:SAIDI}
    {\tt SAIDI} = \frac{\sum U_{i}N_{i}}{\sum N_{i}}
\end{equation}
where $\sum_{\forall i} U_{i}N_{i}$ is {\em total number of customer interruption} durations. ${\tt SAIDI}$ is a measure of the expected duration of interruptions a customer is expected to experience. 
\newline
\item Customer Average Interruption Duration Index
\begin{equation}
    \label{eq:CAIDI}
    {\tt CAIDI} = \frac{\sum U_{i}N_{i}}{\sum_{\forall i} \lambda_i N_{i}} = \frac{\tt SAIDI}{\tt SAIFI}
\end{equation}
${\tt CAIDI}$ is the ratio between ${\tt SAIDI}$ and ${\tt SAIFI}$ and measures the average duration each given customer in the system is expected to experience. 
\end{enumerate}

\subsection{Structure of RELSAD}

\begin{figure}
    \centering
    \includegraphics[width=0.5\textwidth]{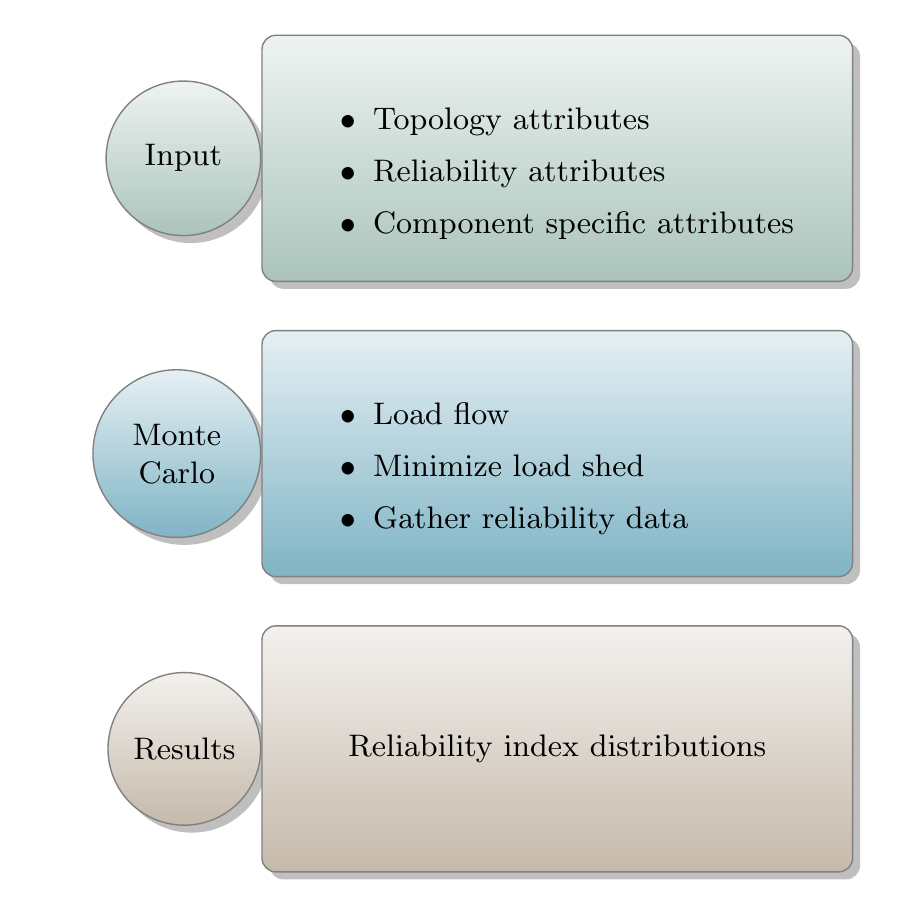}
    \caption{Overview of the RELSAD model. A full description can be found in the software documentation \cite{RELSAD_documentation}.}
  \label{fig:RELSAD_structure}
\end{figure}

In Fig. \ref{fig:RELSAD_structure}, an overview of the RELSAD model is illustrated. RELSAD considers different user-specified attributes such as:

\begin{itemize}
    \item \textbf{Topology attributes:} such as spatial coordinates and connected components.
    \item \textbf{Reliability attributes:} such as failure rates and outage distributions.
    \item \textbf{Component specific attributes:} such as, for a battery, this could be the battery capacity, the inverter capacity, efficiency, and charging limitations. 
\end{itemize}

The core of the software is the simulation where different operations are performed. The general functionality of RELSAD is described in Sec. \ref{sec:core_func_RELSAD} and the microgrid-specific functionalities are described in Sec. \ref{sec:microgrid_implementation}. 

The output or results of the RELSAD software are given as reliability index distributions. Multiple different distributions of the results can be given and the investigated reliability indices in this paper are described in Sec. \ref{sec:rel_indices}. 

RELSAD is constructed in an object-oriented fashion, where the systems and system components are implemented with specific features simulating their real-life behavior. RELSAD is a general tool that can consider any network that operates radially. In RELSAD, a power system - $P_{s}$ is first created before distribution systems - $D_{s}$ and other networks such as Microgrids - $M_{s}$ are created inside the $P_{s}$. After this, the electrical system components can be created and added to associated network layers. The possible electrical components are lines - $l$, buses - $b$, disconnectors - $d$, circuit breakers - $cb$, generation units - $p_{u}$, and batteries - $p_{b}$. In addition, a load - $p_{d}$ can be assigned to each bus. This is described in more detail in the documentation page of the software \cite{RELSAD_documentation}.

\subsection{Core functionality of RELSAD}
\label{sec:core_func_RELSAD}

In this section, the core functionality of RELSAD is described. The reliability assessment of a system is solved through SMCS in RELSAD. To account for the active participation of different technologies such as microgrids, DG, and batteries, a load flow solver is implemented to evaluate the electrical consequence of faults in the system. Through load flow calculations, the behavior of the system during different scenarios can be evaluated. In RELSAD, a Forward-Backward sweep (FBS) approach is implemented as the load flow solver. 
In the FBS approach, the load flow is calculated by updating the power flow through a backward sweep before the voltage magnitudes and angles at the system buses are updated in a forward sweep \cite{haque1996load}. 

Due to the potential island operation of parts of the distribution system with and without the microgrid, power balance needs to be ensured. To achieve this, a simple load shed optimization problem is included. The objective of the load shed optimization problem, seen in eq. \ref{eq:min}, is to minimize the total load shed in the network based on the price of shedding different load types. The price is based on the \textit{Cost of Energy Not Supplied} (CENS). This is subjected to load flow balance and the capacity limitations over the power lines, the load, and the generation in the distribution system:

\begin{align}
    &\underset{P^{s}_{n}}{\text{minimize}}
    \quad \mathcal{P}_s = \sum_{n = 1}^{N_n}   C_{n}\cdot P^{s}_{n} \label{eq:min}\\
    &\text{subject to: } \nonumber \\
    &\begin{aligned}
        \sum_{i=1}^{N_l} \alpha_i \cdot P^{l}_{i} &= \sum_{j=1}^{N_g} \nu_j \cdot P^{g}_{j} - \sum_{k = 1}^{N_n}& \eta_k& \cdot (P^{d}_{k} - P^{s}_{k})\\
        \min P_{j}^{g} &\leq P_{j}^{g} \leq \max P_{j}^{g} &\forall j&=1,\dots,N_{g}\\
        0 &\leq P_{k}^{s} \leq P_{k}^{d}  &\forall k&=1,\dots,N_{n}\\
        \left| P_{i}^{l} \right| &\leq \max P_{i}^{l}  &\forall i&=1,\dots,N_{l}\\
    \end{aligned} \nonumber
\end{align}

Here $C_n$ is the cost of shedding load at node n while $P_{n}^{s}$ is the amount of power shed at node n. $P_{j}^{g}$ is the production from generator $j$. $P_{k}^{d}$ is the load demand at node $k$ while $P_{i}^{l}$ is the power transferred over line $i$. $\alpha_i$ = 1 if line $i$ is the starting point, -1 if line $i$ is the ending point. $\nu_j$ = 1 if there is a production unit at node $j$, otherwise it is 0. $\eta_k$ = 1 if there is a load on node $k$, otherwise it is 0.

\subsubsection{\textbf{Incremental procedure of fault handling in RELSAD}}

The SMCS in RELSAD is constructed to have a user-chosen increment. This can, for example, be a second, a minute, an hour, or a day. 
The incremental procedure of the fault handling model implemented in RELSAD is illustrated in Algorithm \ref{alg:procedure}. After a $P_{s}$ is created with associated systems and components, the incremental procedure will set the load and generation at the system buses for the current time increment. Then the failure status of each component is calculated based on random sampling and the failure rate of the different components. If any component is in a failed state, the network with the failed components will be divided into sub-systems, and a load flow and load shedding optimization problem will be solved for each sub-system. The load flow and load shedding optimization problem are included to calculate the electrical consequence of a fault when active components are present. When this is performed, the historical variables of the components are updated. The historical variables can then further be used to evaluate the reliability of the $P_{s}$, the different networks in the power system, and the individual load points. An example of this is provided in the documentation page for the software \cite{RELSAD_documentation}.  
The rest of this section describes the different procedures of the reliability evaluation in greater detail. 

\begin{algorithm}
    \SetAlgoLined
     Set bus $p_d$ and $p_u$ for the current time increment\;
     Draw component fail status\;
     \If{Failure in $P_s$ }{
         Find sub-systems\;
         \ForEach{sub-system}{
            Update $p_b$ and $p_{EV}$ demand (charge or discharge rate, discharge of EVs if V2G is activated)\;
            Run load flow\;
            Run load shedding optimization problem\;
         }
         Update history variables\;
     }
    \caption{Increment procedure}
    \label{alg:procedure}
\end{algorithm}

\subsection{Microgrid reliability}
\label{sec:microgrid_implementation}

Reliability evaluation will depend on the operation mode of the microgrid controller. If the microgrid is in island mode, it will function as an independent system. Whereas, if the microgrid is connected to the overlying distribution system during the outage period, it will be a part of the distribution system and the systems need to interact. In \cite{wang2013new}, new metrics for the reliability of microgrids in island mode are proposed. 
In this study, the microgrid is designed to primarily be connected to the distribution system. The island mode of the microgrid is only achieved during outages in one of the systems. This will then make the regular reliability indices well-suited for evaluating the reliability of the microgrid. 

To achieve the right commands and the interaction between the microgrid and the distribution system, a microgrid controller is implemented. The microgrid controller will in this paper focus on performing the correct fault handling; the remaining features of the controller are simplified as they are of minor importance to this study.
Here, the possibilities for operating modes can be extensive, but for this study, three different operating modes were implemented:

\begin{enumerate}
    \item \textbf{No support mode (island mode)}--–During failures in the system, the microgrid will work in island mode during the down time of the failure.
    \item \textbf{Full support mode}-–-After the sectioning time, the microgrid will reconnect to the distribution network (if possible) and contribute to the operation of the distribution network.
    \item \textbf{Limited support mode}-–-After the sectioning time, the microgrid will reconnect to the distribution network (if possible) but will only contribute with limited power from the battery so that the microgrid can ensure self-sufficiency for a limited period of time. This means that the battery will store enough power to operate the microgrid for four peak hours. The battery will, in this scenario, function as a backup source, and the SoC will not be drawn when a failure occurs but will start with max SoC.
\end{enumerate}
The microgrid controller was implemented to focus on performing the correct fault handling; the remaining features are simplified.


Algorithm \ref{alg:Microgrid_procedure} illustrates the controller procedure for fault handling in the microgrid. First, the \textit{sectioning time} of the $cb$ is updated if needed. The sectioning time is the time it takes from a failure occurring on a line until the fault is located and disconnected. The sectioning time of the $cb$ is updated in case the system was in a sectioning time in the previous time step. 
Then, if a new outage happens in one of the systems, $D_s$, and $M_s$, the circuit breaker of the $M_s$ will open. 
In the $M_s$, all lines will be checked for failures. If the line containing the circuit breaker has failed, then the circuit breaker should remain open regardless of the $M_s$ mode. If not, the algorithm will check in which mode the $M_s$ should operate. 
If the $M_s$ is supposed to operate in island mode during faults in the $D_s$, the procedure will check for failed lines in the $D_s$. In case of no failed lines in the $D_s$, the failed sections in the $M_s$ are disconnected and the circuit breaker is closed. If the microgrid is in a supportive mode, where it can be islanded with parts of the $D_s$ buses, the failed sections in the $M_s$ are disconnected before the circuit breaker is closed. 

This will result in the $M_s$ being in one of three states during an outage in the $P_s$: 1) in island mode, 2) in island mode with parts of the $D_s$, or 3) connected to the $D_s$ with the possibility of supply from the overlying network. 

\begin{algorithm}
    \SetAlgoLined
    \LinesNumbered
    Update sectioning time\;
    \If{$cb$ is open and sectioning time $\leq$ 0 or $l$ recovered from failure} 
    { 
    Check $l$ in $M_s$ for failure\; 
    \If{$l$ with $cb$ not failed}{
        \uIf{$M_s$ in island (survival) mode}{
            \If{No failed $l$ in $D_s$}{
                Disconnect failed sections\;
                Close $cb$\;
            }
        }
        \Else{
            Disconnect failed sections\;
            Close $cb$\;
        }
    }
    }
    \caption{Microgrid controller procedure}
    \label{alg:Microgrid_procedure}
\end{algorithm}

\subsection{Statistical analysis}



\subsubsection{Statistical difference}

Statistical analysis is important to confirm the hypotheses made by researchers. A statistical test checking for equal means can be performed to quantify whether the result data sets from various scenarios differ. There are two classes of hypothesis tests, namely 1) \textit{parametric} \cite{t-test} and 2) \textit{non-parametric} \cite{KS}. Both classes are bounded by assumptions about the distribution of the result data sets. Parametric tests are restricted to normally distributed data sets with equal variance, while non-parametric tests have looser restrictions. For this reason, parametric tests have stronger statistical power, making them the preferred choice if the data set satisfies its assumptions. To decide which class of tests to use, an evaluation of the distribution of the results data set must be carried out. 

The evaluation of normality can be done by performing an Anderson-Darling (AD) test. The AD test is used to establish if a data set comes from a population with a specific statistical distribution \cite{Stephens1974}. The outcome of the test is measured by the deviation \textit{$A^{2}$}, between the number of samples and a weighted logarithmic expression of the data in the data set. An increasing deviation between \textit{$A^{2}$} and zero decreases the probability of the data set being of a specific statistical distribution.

If the AD test for normality fails, non-parametric tests must be used to check for equal means. In this paper, the Kolmogorov-Smirnov (KS) \cite{KS} test is used for this purpose. The KS test quantifies the distance between two data sets, indicating if the data sets are drawn from the same population. 

\subsubsection{Sensitivity analysis}

Sensitivity analysis is important since it provides a mapping of the behavior of the model and illustrates how the results are affected by various parameters. 

Factorial design is a more advanced form of sensitivity analysis 
used to investigate and understand the effect of independent input variables on a dependent output variable \cite{Statisticbook}. One of the key advantages of factorial design compared to a conventional sensitivity analysis---that only studies the effect of one variable at a time---is the ability to quantify the influence the variables have on the output variable \textit{and} on the other investigated variables. However, this drastically increases the number of combinations to investigate; which will lead to more simulation time, for example, a factorial design test with three factors that consider two levels each leads to a total of $2^{3} = 8$ combinations.

\section{Reliability test system}
\label{CaseStudy}

In this section, the reliability test system with the procedures for data gathering, and the case study are presented. First, we introduce the IEEE 33-bus system, which is the distribution network used for reliability assessment in this paper. Secondly, we present the microgrid and the microgrid parameters that are studied. Finally, the case scenarios are outlined. 

\subsection{Procedures for developing the test network}
\label{ProcedureTestNetwork}

\begin{figure*}[tb]
    \centering
    \includegraphics[width=\textwidth]{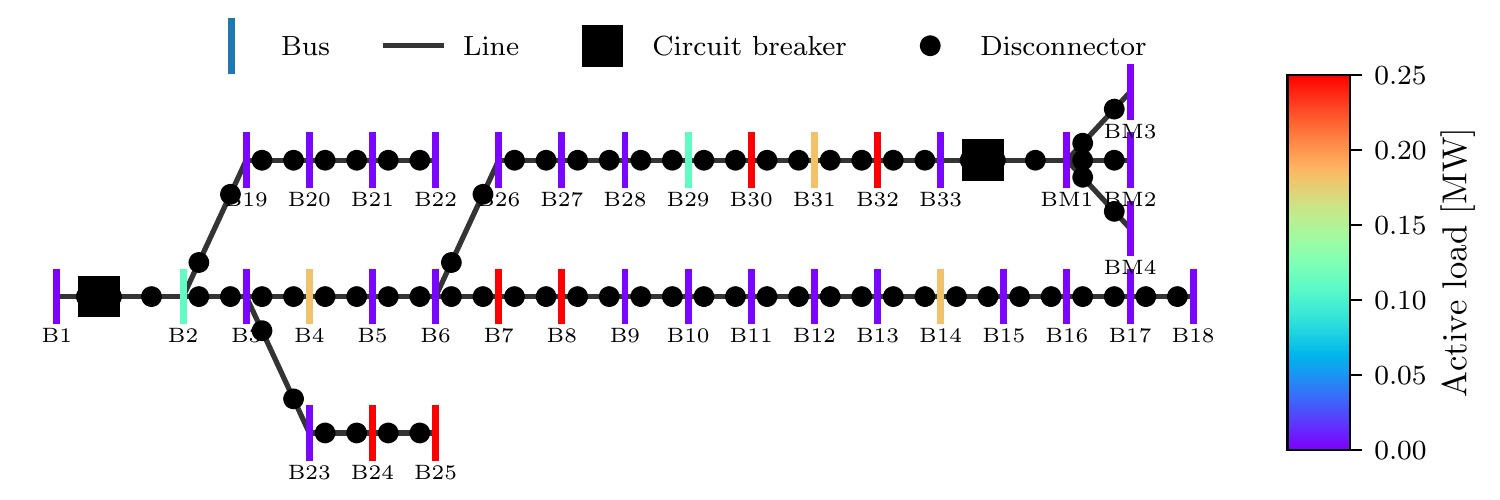}
  \caption{The system topology. The microgrid is placed on B33. The plot illustrates a heat map of the mean load at each bus in the system.}
  \label{fig:SystemTopology}
\end{figure*}

The method is demonstrated on the IEEE 33-bus system seen in Fig. \ref{fig:SystemTopology}. The IEEE 33-bus system is chosen since it is a commonly used test network, and therefore, makes it easier for evaluation and replication. 
The network is operated fully radially without any backup connections, and possible islanded operation with the microgrid will be investigated. To account for a more realistic distribution network, the customers in the network with their demand profiles are distributed to map a real network. To make a dynamic demand profile of non-constant load, load profiles generated based on the FASIT requirement specification with hourly time variation of the load are used \cite{FASIT}. In Fig. \ref{fig:SystemTopology}, the mean load at each bus is illustrated through a heat map. The maximum load in the distribution system is scaled to approximately double the size of the traditional load values in the IEEE 33-bus system (see \cite{baran1989network} for the original network). The distribution system is constructed to include different types of loads, such as households, farms, industry, trade, and office buildings, to build a more dynamic customer profile in the system and include the priority of the loads. The weather data is collected from a location in eastern Norway. The load data can be seen in \cite{RELSAD}.

In this study, only failures on the lines in the network will be addressed. The reliability data used in the study is collected from yearly reliability statistics from the Norwegian distribution systems \cite{statnett_2018}. The average failure rate of the lines is 0.07 failures/year/km. The lines in the test system are based on the original lines in the IEEE 33-bus network \cite{baran1989network}. However, the length of the lines is calculated based on the impedance of the lines and gives lines of different lengths in the system. 

\begin{figure}[h]
    \centering
    \includegraphics[width=0.5\textwidth]{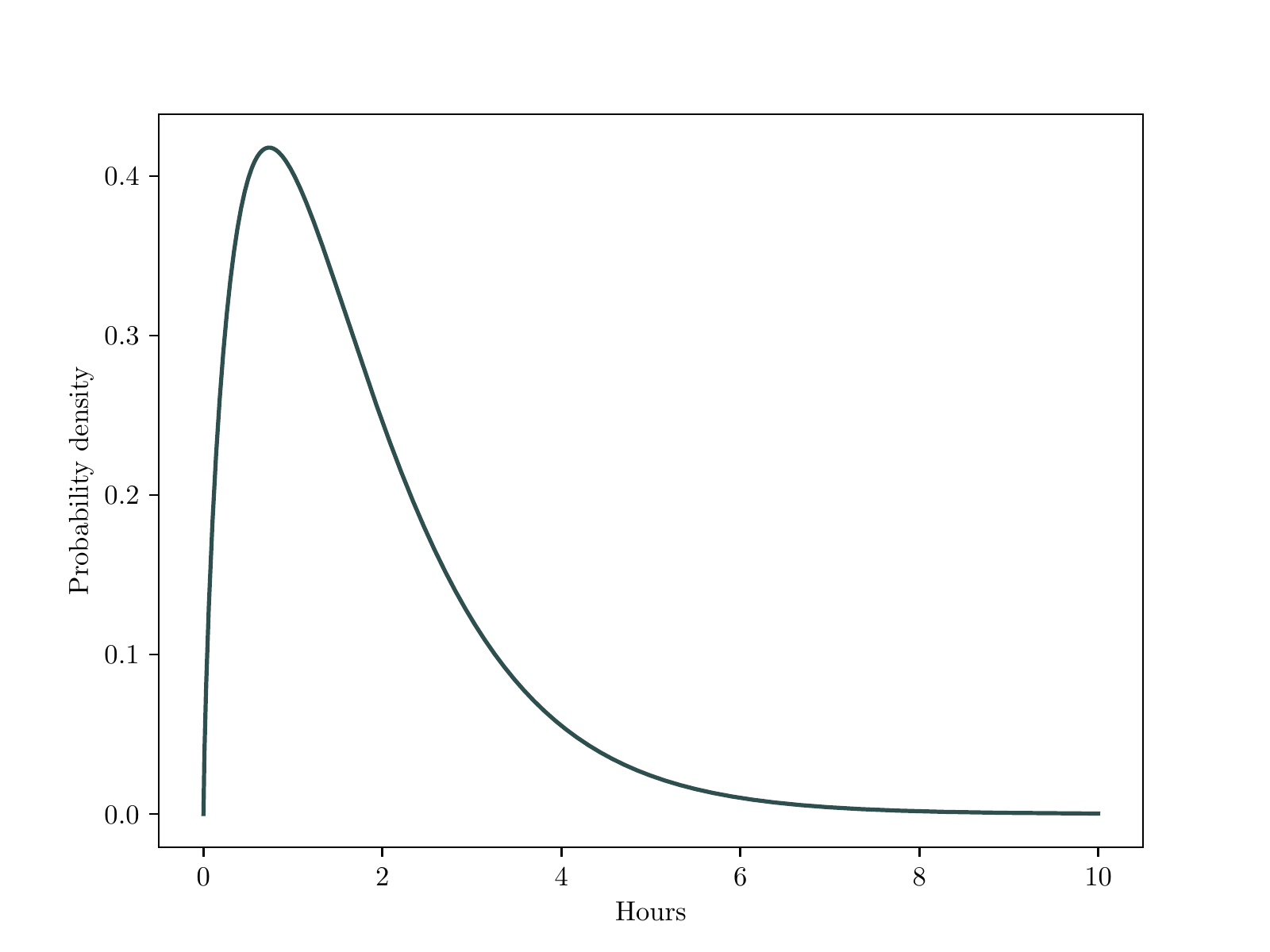}
  \caption{Gamma distribution of the repair time of the lines in the system.}
  \label{fig:outagetime_dist}
\end{figure}

In the statistics from the Norwegian DSOs, the repair time of the line is given as a percentage of the amount of time a failure lasts for a given time period. On average, the repair time will be less than two hours, 67\% of the time. For creating various line repair times, the repair time follows a gamma distribution as seen in Fig. \ref{fig:outagetime_dist} where the line repair time will be two hours or less, 67\% of the time.

\subsection{Microgrid}

The microgrid is placed on bus 33 at the end of the radial arm in the distribution system.
The microgrid contains wind power, solar power, a battery, and some load. The specifications of the microgrid can be seen in Tab. \ref{tab:MicrogridSpec}. The wind and solar power profiles are generated based on weather data from the same location in Norway with hourly variation and follow the methods presented in \cite{RELSAD_documentation}. 

The microgrid is operated in grid-connection mode most of the time, but if a line failure occurs in the distribution system or the microgrid, the microgrid will shift to island operation.

\begin{table}[h]
    \caption{Microgrid specifications}
    \center
\begin{tabular}{|c|c|} 
     \hline
     \textbf{Component} & \textbf{Specification} \\ \hline
     Battery &  Max capacity: 1 MWh \\ 
     \cline{2-2}
     & Inverter capacity: 500 kW \\ 
     \cline{2-2}
      & Efficiency: 0.95 \\ 
      \cline{2-2}
      & Min SOC: 0.1 \\ \hline
     Wind \& Solar power 
     & Max power: $\sim$ 3.5 MW \\\hline
     Load & Peak load $\sim$  200 kWh  \\ \hline
\end{tabular}
\label{tab:MicrogridSpec}
\end{table}

\subsubsection{\bf Battery strategy}

In multiple cases, batteries in the power system will sell and buy power in a power market or flexibility market. If the battery participates in such a market, the stored energy in the battery will vary over time, and might not be full when a failure occurs. 
In this study, we have considered variations of stored energy in the battery by calculating the state of charge (SoC) level. The SoC level is assumed to follow a uniform distribution since we do not have any information about the market strategy. This will create a more realistic case where the battery is not necessarily full when a failure occurs.

\subsubsection{\textbf{Description of the different scenarios}}

The case study investigates three different scenarios, which are based on three different operating modes for the microgrid, namely: 1) \textit{no support mode}, 2) \textit{full support mode}, and 3) \textit{limited support mode}, as described in Sec \ref{sec:microgrid_implementation}. The reliability for both the distribution system and the microgrid will be measured in ${\tt ENS}$.

\section{Results and discussion}
\label{Results}

This section presents the results of the three conducted reliability studies. First, the results for the three described scenarios are provided from both the distribution system and microgrid perspective. Second, we present a sensitivity analysis of how the battery capacity, line repair time, and line failure rate affect the ${\tt ENS}$ of both the distribution system and the microgrid. Finally, the investigation of how the microgrid location affects the ${\tt ENS}$ of the distribution system is detailed. Steady-state conditions are assumed for all the reliability studies. The simulations converge after approximately 3,000 iterations, and the study has been performed with 5,000 iterations.

\subsection{Microgrid scenario study}

\subsubsection{\textbf{The distribution system perspective}}

\begin{figure}[h]
    \centering
    \includegraphics[width=0.5\textwidth]{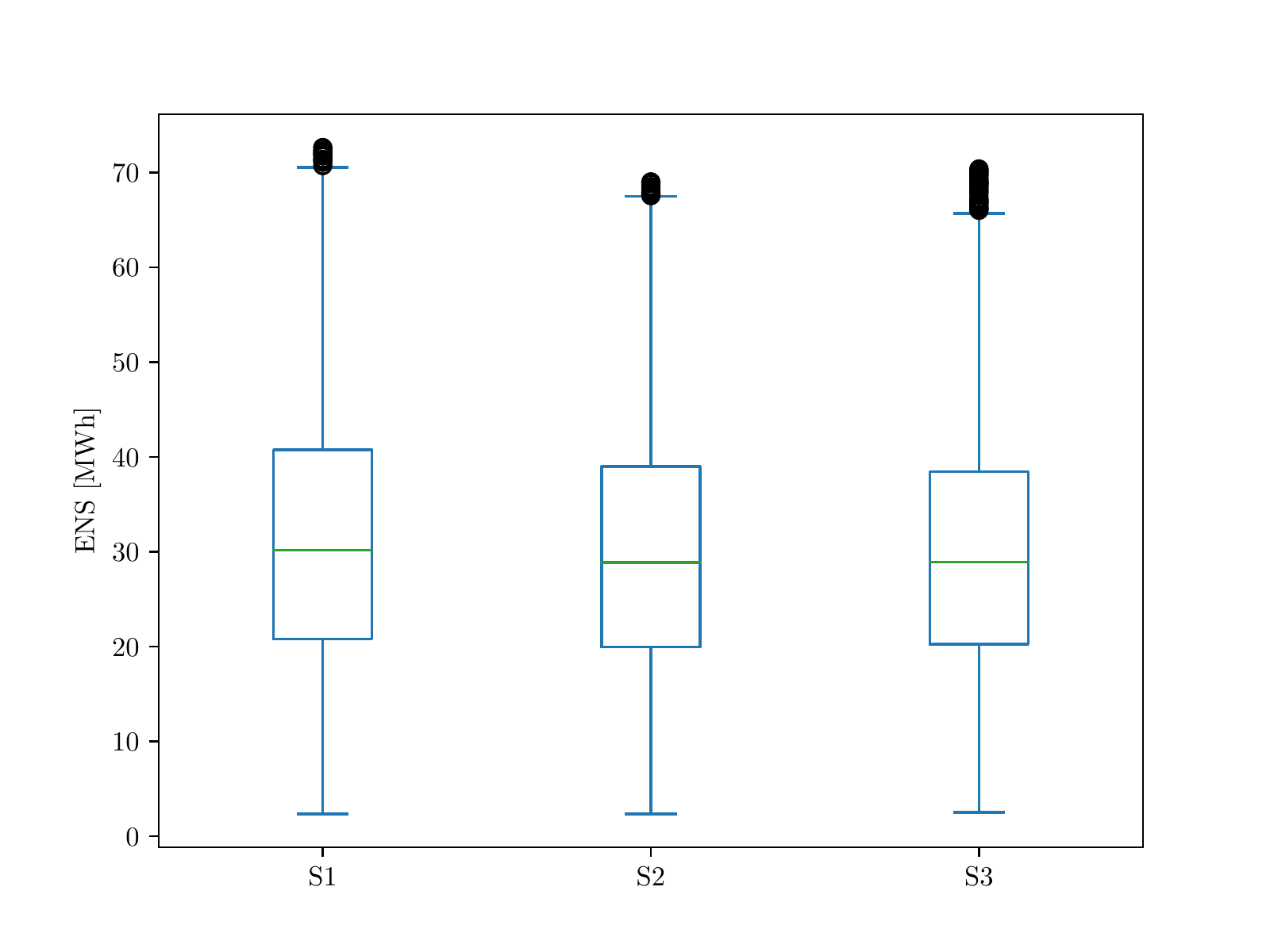}
  \caption{Box plot of the frequency distribution of energy not supplied in the distribution system for the three scenarios.}
  \label{fig:Boxplot3scenariosD}
\end{figure}

In Fig. \ref{fig:Boxplot3scenariosD}, a box plot of the ${\tt ENS}$ for the distribution system is illustrated. It can be observed from the box plot that there is a slight decrease in ${\tt ENS}$ for Scenarios 2 and 3 where the microgrid supports the distribution system during faults, compared to Scenario 1 where the microgrid operates in island mode during faults. 
Comparing Scenarios 2 and 3, there are only minor differences. 

The result is more apparent in Tab \ref{tab:ReliabilityIndicesD} where the mean values of the discussed reliability indices are presented for the three scenarios. Overall, the mean ${\tt ENS}$ decreases by 4.27\% for Scenario 2 and 4.62\% for Scenario 3 compared to Scenario 1. 
This effect could be increased by having multiple generation sources or microgrids spread around the network. Since the microgrid is located in one place in the microgrid, there will be several cases where the microgrid is unable to support the grid since the isolated part is not connected to the microgrid. In addition, since the microgrid prioritizes its own load, the distribution system will get less support. 

\begin{figure}[h]
    \centering
    \includegraphics[width=0.5\textwidth]{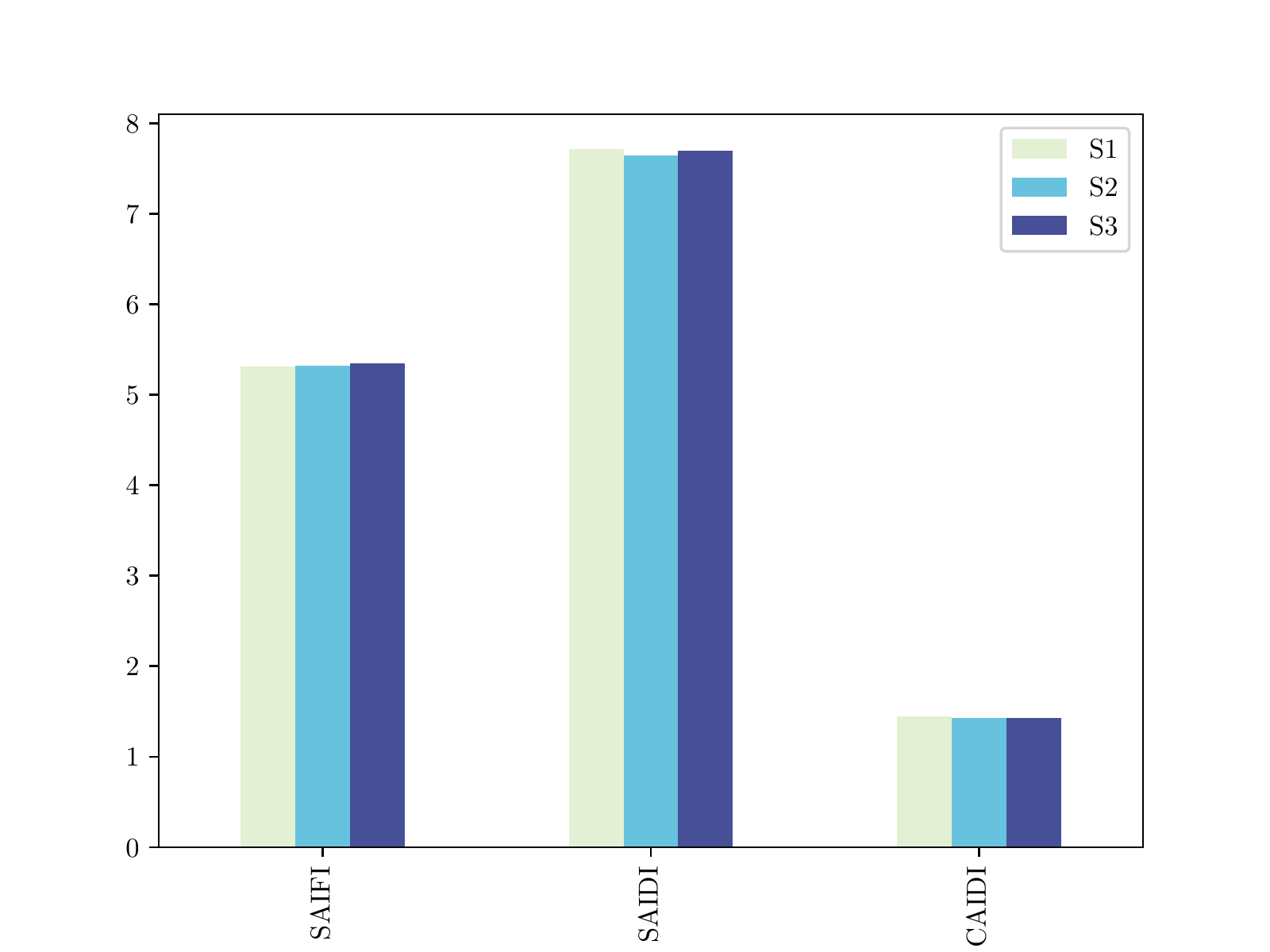}
  \caption{Bar plot of the mean values of the reliability indices for the distribution system. (SAIFI is given in frequency of disturbances while SAIDI and CAIDI are given in hours).}
  \label{fig:reliability_indices_dist}
\end{figure}

\begin{table}[h]
    \caption{Mean values of the reliability indices for the distribution system}
    \center
\begin{tabular}{c|ccc} 
     \hline
     & \textbf{S1} & \textbf{S2} & \textbf{S3}\\ \hline
     \textbf{ENS [MWh]} & 31.7518 & 30.3975 & 30.2834 \\ 
     \textbf{SAIFI [freq.]} & 5.3566 & 5.3630 & 5.3925 \\ 
     \textbf{SAIDI [h]} & 7.7945 & 7.7261 & 7.7745 \\ 
     \textbf{CAIDI [h]} & 1.4551 & 1.4406 & 1.4417 \\ 
     \hline
\end{tabular}
\label{tab:ReliabilityIndicesD}
\end{table}

Fig. \ref{fig:reliability_indices_dist} and Tab. \ref{tab:ReliabilityIndicesD} display the other investigated reliability indices (SAIFI, SAIDI, and CAIDI) for the distribution system. By analyzing these indices, some small differences between the different scenarios can be observed. The down time for the system decreases with support from the microgrid. The reason for this is that some load points will experience a decreased period of shedding. However, the failure frequency will increase. In most cases, the microgrid generation and battery are unable to preserve supply during the entire outage period, resulting in some load points experiencing a \textit{new} outage period when the generation is low or the battery is empty. In addition, the down time for Scenario 3 is slightly higher than for Scenario 2. This is a result of the microgrid storing power for its own load, leading to less power for the distribution system.

Statistical testing was performed on the results from the three different scenarios. First, the AD test was used to investigate if the results are normally distributed. As seen in Tab. \ref{tab:Stat_test_D_AD}, the normality test failed, indicating that none of the results are normally distributed. The test statistics, $A^2$, are very high, indicating with high certainty that the result is not normally distributed.

The non-parametric test was then performed to investigate if the result originates from the same population. A significance level of 5\% was chosen, meaning that the null hypothesis, which is that the result is taken from the same population, is rejected if the p-value of the test is lower than 0.05. Tab. \ref{tab:Stat_test_D_AD} shows that Scenario 1 is significantly different from Scenarios 2 and 3. However, Scenarios 2 and 3 are not significantly different, meaning they are from the same population. This result indicates that there is a significant difference between receiving support from the microgrid during unintentional outages in the distribution system compared to no support. However, the actual network impact might vary.


\begin{table}[h]
    \caption{Statistical test of the three scenarios in the distribution system}
    \center
\begin{tabular}{c|c||c|c} 
     \hline
     \textbf{Scenario} & \textbf{AD test} &\textbf{Scenarios} & \textbf{KS test}\\
     & $A^{2}$ & & p-value \\ \hline
    
    \textbf{S1} & 22.34 &\textbf{S1 vs. S2}& 0.00\\ 
    \textbf{S2} & 20.68 &\textbf{S1 vs. S3}& 0.00\\ 
    \textbf{S3} & 18.42 &\textbf{S2 vs. S3}& 0.53\\ \hline

\end{tabular}
\label{tab:Stat_test_D_AD}
\end{table}

\subsubsection{\textbf{The microgrid perspective}}

\begin{figure}[h]
    \centering
    \includegraphics[width=0.5\textwidth]{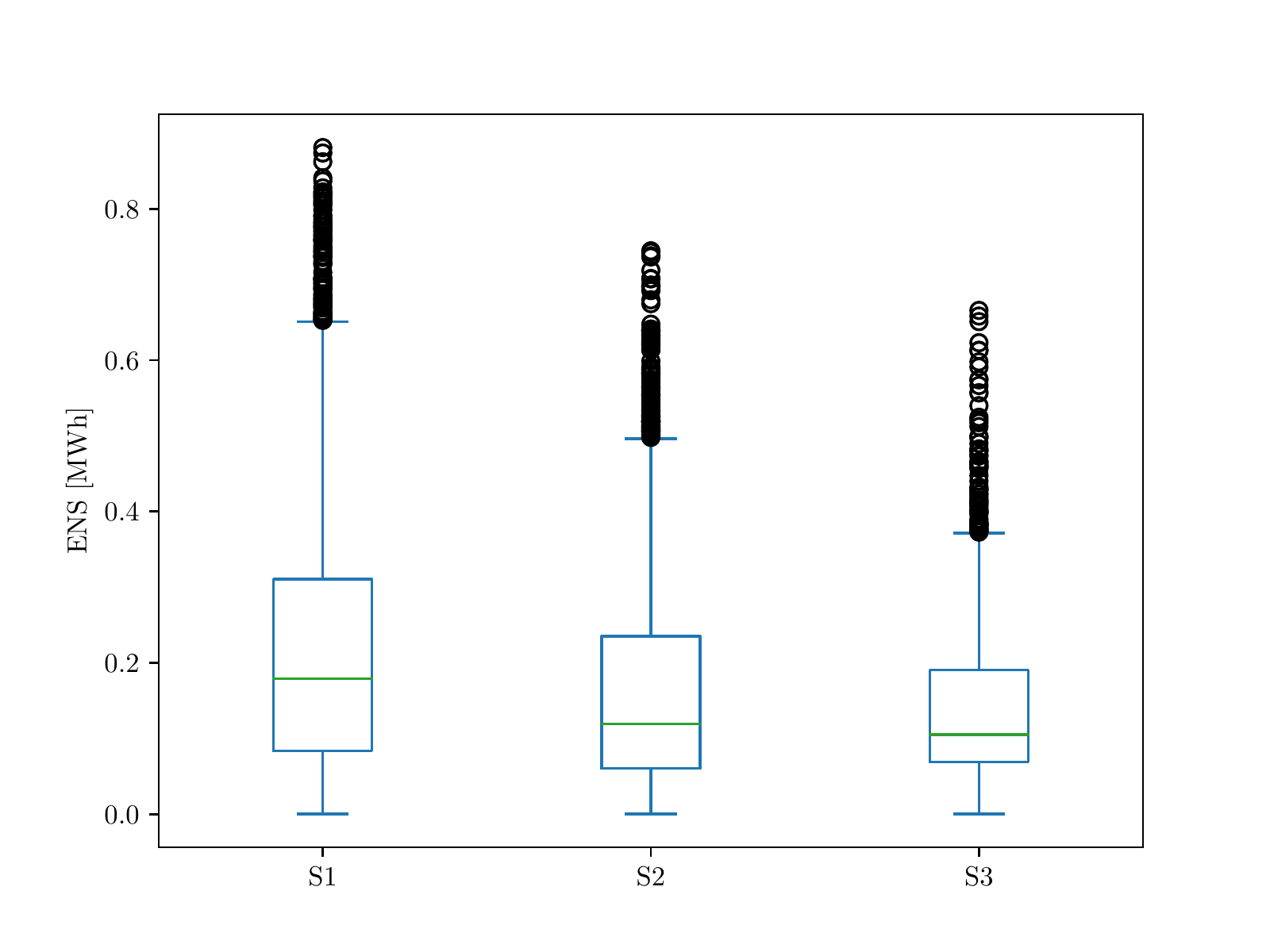}
  \caption{Box plot of the frequency distribution of energy not supplied in the microgrid for the three scenarios.}
  \label{fig:Boxplot3scenariosM}
\end{figure}

The amount of ${\tt ENS}$ from the microgrid perspective for the different scenarios can be seen in Fig. \ref{fig:Boxplot3scenariosM}.
The differences between the scenarios are clearer when investigating them from the perspective of the microgrid. Here, Scenario 1 will result in most shedding, since the microgrid has to survive in island mode during faults in the distribution system without a backup supply. Scenario 2 is somewhat better since the microgrid will reconnect to the distribution system. For situations where the microgrid is in the same sub-system as the main feeder of the distribution system, the microgrid will be ensured supply. Scenario 3 is the case that will give a very small ${\tt ENS}$. The shedding and outliers in the box plot are a result of faults inside the microgrid, long repair time on some failures, and some \textit{high impact low probability} cases. These cases are the result of two failures happening in close approximation, leading to a discharged battery before the last fault is removed. 

\begin{figure}[h]
    \centering
    \includegraphics[width=0.5\textwidth]{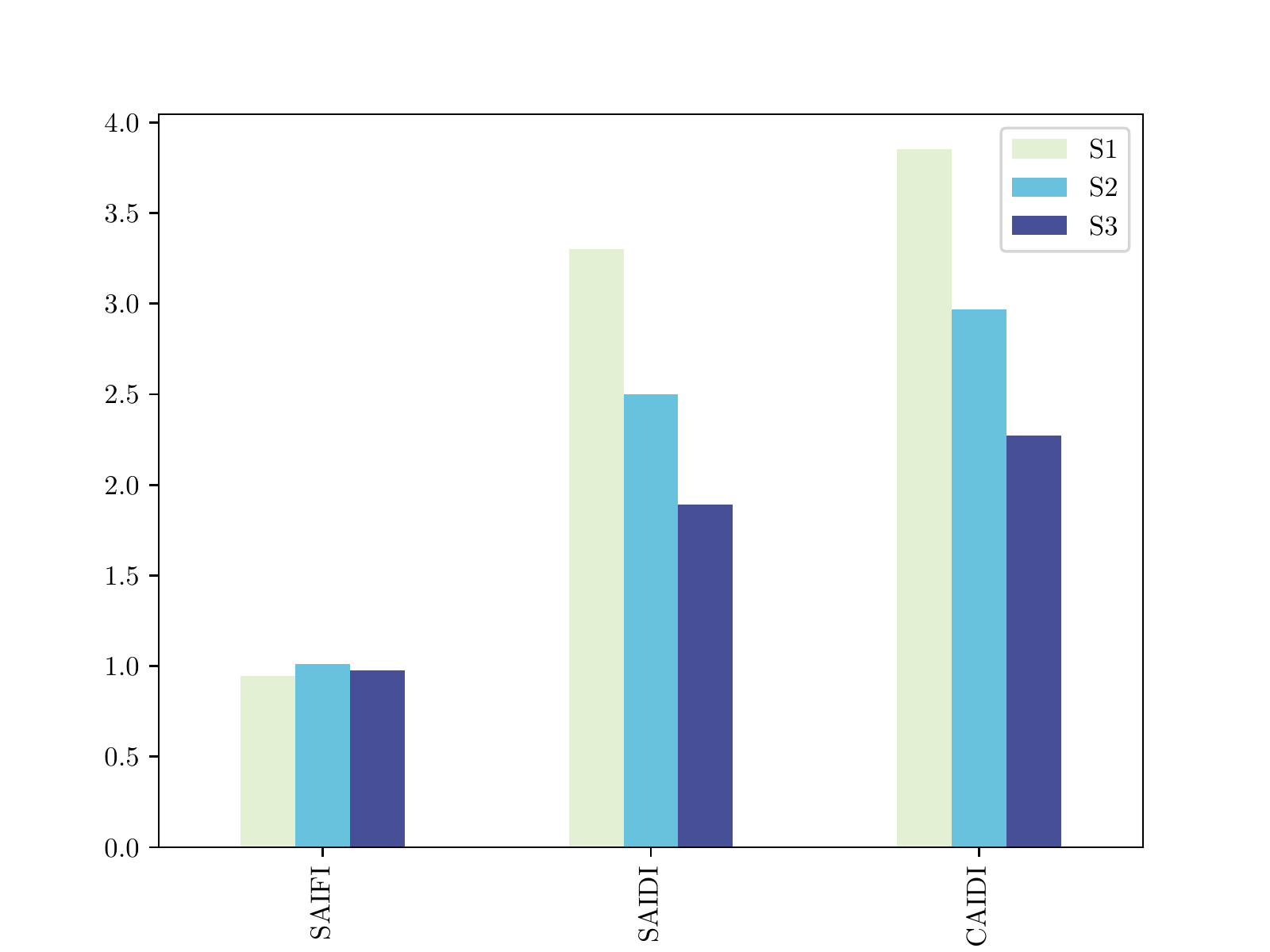}
  \caption{Bar plot of the mean values of the reliability indices for the microgrid. (SAIFI is given in frequency of disturbances while SAIDI and CAIDI is given in hours).}
  \label{fig:reliability_indices_microgrid}
\end{figure}

The mean values of the reliability indices for the microgrid can be seen in Fig. \ref{fig:reliability_indices_microgrid} and Tab. \ref{tab:ReliabilityIndicesM}. The indices indicate that there is a considerable decrease in the interruption duration for Scenario 3 compared to the others. Scenario 3 also leads to a lower interruption frequency. The increase in interruption frequency for Scenario 2 is caused by low generation and an empty battery that is not sufficient to supply the microgrid load resulting in the microgrid load experiencing additional interruption. 
The ${\tt ENS}$ for Scenario 3 decreased by 69.32\%
compared to Scenario 1, whereas the ${\tt ENS}$ for Scenario 2 decreased by 26.63\%.

\begin{table}[h]
    \caption{Mean values of the reliability indices for the microgrid}
    \center
\begin{tabular}{c|ccc} 
     \hline
     & \textbf{S1} & \textbf{S2} & \textbf{S3}\\ \hline
     \textbf{ENS [MWh]} & 0.1258 & 0.0923 &  0.0386 \\ 
     \textbf{SAIFI [freq.]} & 0.5345 & 0.5556 & 0.2554 \\ 
     \textbf{SAIDI [h]} & 1.8814 & 1.3890 & 0.5122 \\ 
     \textbf{CAIDI [h]} & 3.5201 & 2.4999 & 2.0059 \\ 
     \hline
\end{tabular}
\label{tab:ReliabilityIndicesM}
\end{table}

The same statistical tests were performed on the reliability result for the microgrid. The result can be seen in Tab. \ref{tab:Stat_test_M_AD}. None of the results passed the normality test, which can be expected based on the formation of the box plots. The same non-parametric test was performed with a significance level of 5\%. The outcome shows that all the scenario results are significantly different from each other. The result highlights the importance of backup supply for the microgrid, either through the distribution system or through the microgrid sources. Since all the scenarios are significantly different, Scenario 3 is the best scenario from the microgrid perspective when measuring ${\tt ENS}$.



\begin{table}[h]
    \caption{Statistical test of the three scenarios in the microgrid}
    \center
\begin{tabular}{c|c||c|c} 
     \hline
     \textbf{Scenario}& \textbf{AD test} & \textbf{Scenarios}& \textbf{KS test}\\
     & $A^{2}$  & & p-value\\ \hline
    
    \textbf{S1} & 452.52 &\textbf{S1 vs. S2}  & 0.00\\ 
    \textbf{S2} & 509.56 &\textbf{S1 vs. S3} & 0.00\\ 
    \textbf{S3} & 993.91 &\textbf{S2 vs. S3}  & 0.00\\ \hline

\end{tabular}
\label{tab:Stat_test_M_AD}
\end{table}

\subsection{Sensitivity analysis}

Since there was no significant difference between the results from Scenario 2 and 3 for the distribution system and Scenario 3 gave the best overall result for the microgrid, this is the most appropriate case to investigate further. A full factorial design study was performed on Scenario 3, which was chosen based on the results obtained from the reliability assessment. The investigated parameters and parameter values can be seen in Tab. \ref{tab:factorial}. The repair time of the lines still follows a gamma distribution as described, and the parameters indicate that the repair time will be less than the given parameter value for 67\% of the time. No lower limit values for the battery capacity were chosen since the battery in Scenario 3 is dimensioned for the peak load in the microgrid. Scenario 3, as it is illustrated in this study, is not applicable if the battery capacity is decreased. 



\begin{table}[h]
    \caption{Parameter values used in the factorial design study}
    \center
\begin{tabular}{|c|c|c|} 
     \hline
     & &  \textbf{Name} \\
     \cline{3-3}
     \textbf{Battery} & 1 & ${\tt Bat_{1}}$\\
     \textbf{[MWh]}& 2 & ${\tt Bat_{2}}$ \\ \hline
     &  & \textbf{Name}\\ 
     \cline{3-3}
     & 1 &  ${\tt r_{1}}$\\
     \textbf{Repair time} & 2 & ${\tt r_{2}}$\\ 
     \textbf{[h]}& 3 & ${\tt r_{3}}$ \\ \hline
     & & \textbf{Name}\\ 
     \cline{3-3}
     & 0.05 & ${\tt \lambda_{1}}$\\
     \textbf{Failure rate} & 0.07 & ${\tt \lambda_{2}}$\\
     \textbf{[failures/year]}& 0.09 & ${\tt \lambda_{3}}$\\ \hline

\end{tabular}
\label{tab:factorial}
\end{table}

\begin{figure*}[tb]
    \centering
    \includegraphics[width=\textwidth]{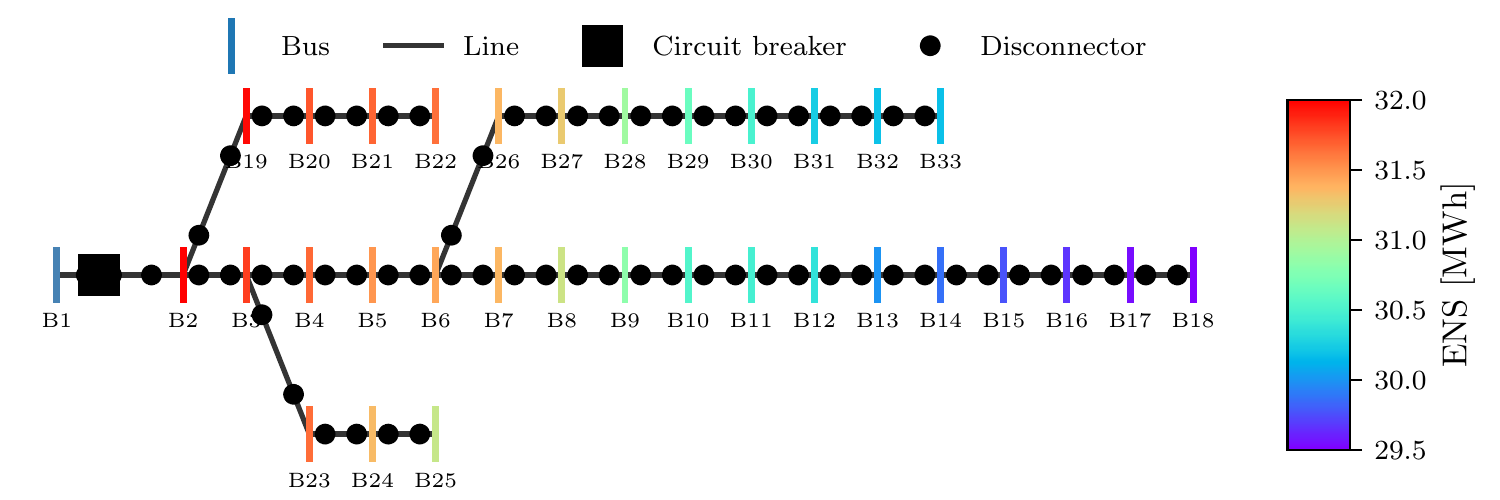}
  \caption{Heat map showing ${\tt ENS}$ in the distribution system for different placements of the microgrid.}
  \label{fig:heatmap}
\end{figure*}

\begin{figure}[h]
    \centering
    \includegraphics[width=0.5\textwidth]{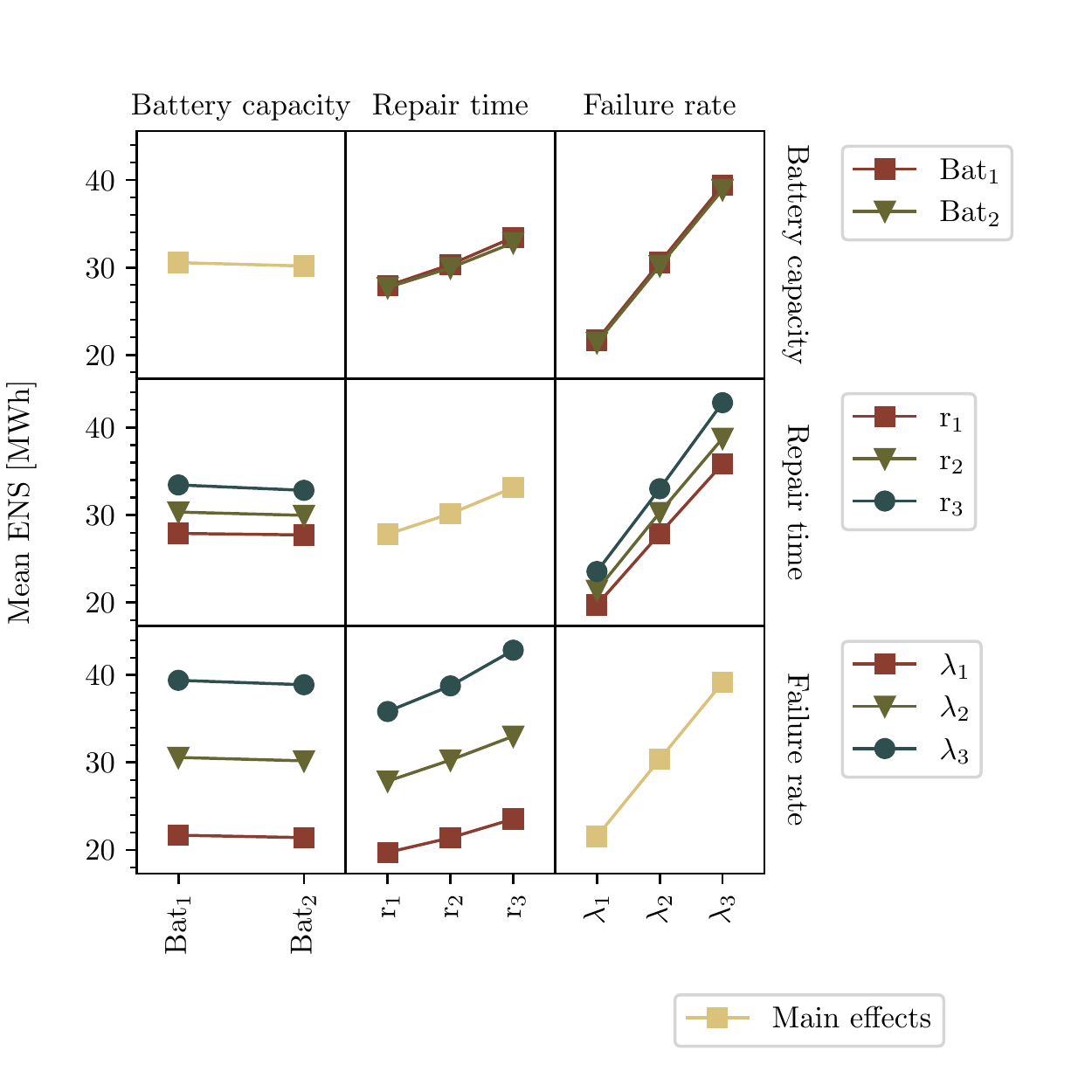}
  \caption{Interaction plot for the distribution network of the mean $\tt ENS$ for the battery, outage time, and failure rate parameters.}
  \label{fig:InteractionDistribution}
\end{figure}

\begin{figure}[h]
    \centering
    \includegraphics[width=0.5\textwidth]{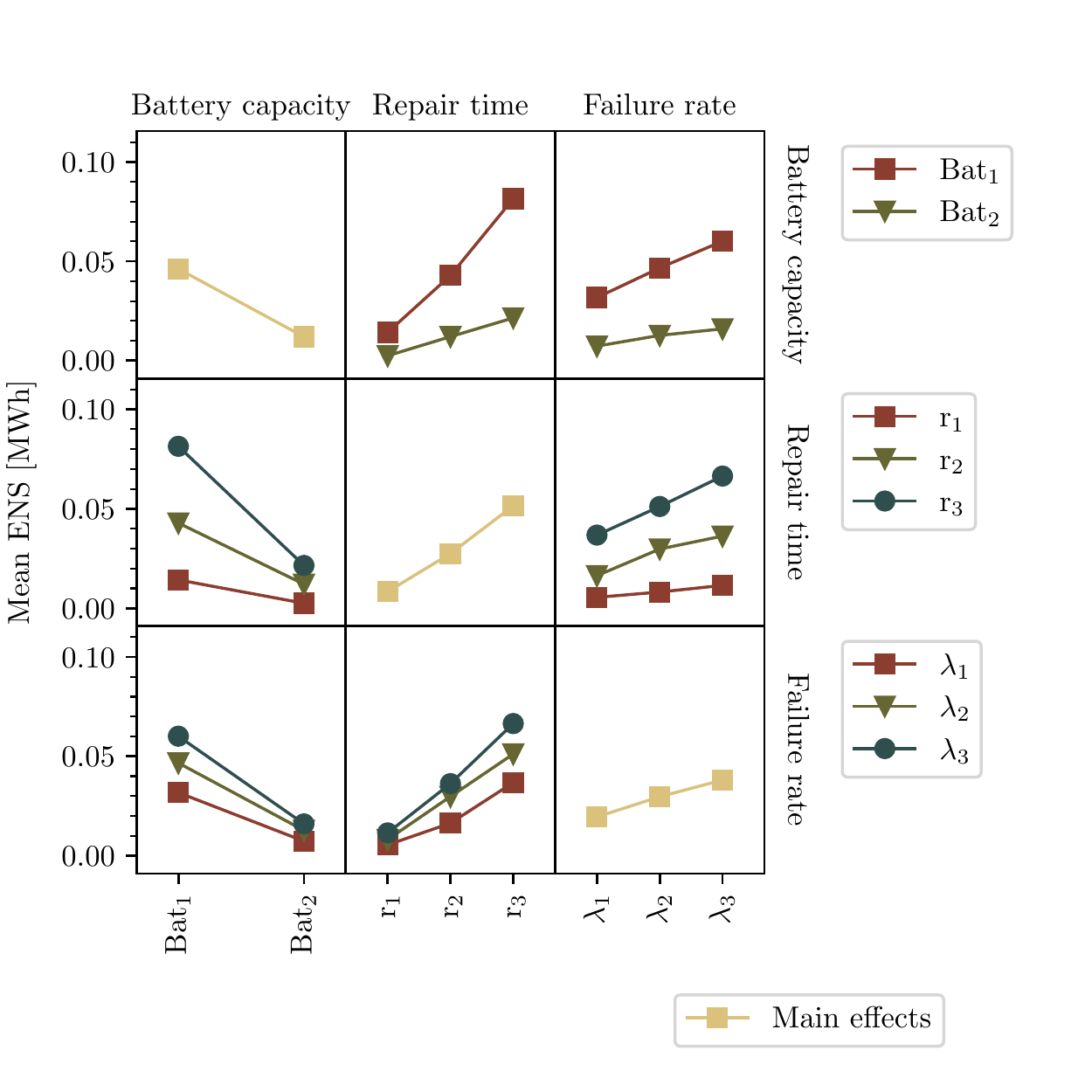}
  \caption{Interaction plot for the microgrid of the mean $\tt ENS$ for the battery, outage time, and failure rate parameters.}
  \label{fig:InteractionMicrogrid}
\end{figure}

The results from the factorial design study are illustrated in Fig. \ref{fig:InteractionDistribution} for the distribution system and in Fig.\ref{fig:InteractionMicrogrid} for the microgrid. The plots indicate both the effect each parameter has on the ${\tt ENS}$ in the system and how the parameters affect each other. 

When investigating the result for the distribution system (Fig. \ref{fig:InteractionDistribution}), the failure rate of the lines gives the largest contribution to ${\tt ENS}$ in the system. Increased battery capacity, however, does not affect the ${\tt ENS}$. This effect is a consequence of the battery being placed at one location in the network, and which is not able to contribute to all possible outage scenarios in the network since it will be isolated from the fault. Unlike the battery, the line failure rate and repair time affect all the lines in the system. 



Investigating the results from the microgrid perspective (Fig. \ref{fig:InteractionMicrogrid}), the battery capacity has a more important role compared to the distribution system. There is a clear interaction effect between the battery capacity and the repair time. 
The ${\tt ENS}$ in the microgrid is more sensitive to variations in repair time when the battery capacity is low. This is a consequence of the microgrid being able to supply the microgrid load for more hours during special \textit{high impact low probability} events, as discussed.
A smaller but similar interaction effect is seen between the battery capacity and the failure rate. When the battery capacity is small, the microgrid is more vulnerable to failures. 


\subsection{Microgrid location study}

We have also investigated the impact of the microgrid placement on the ${\tt ENS}$ in the distribution system. Simulations varying the microgrid location to cover all the distribution system buses were conducted. The results are presented in Fig. \ref{fig:heatmap}. Each bus is colored based on the ${\tt ENS}$ of the distribution system with the microgrid located at the respective bus. The results indicate that it is beneficial to place the microgrid at the end of long radials. In this study, it will be at B33 in the distribution system.

\subsection{Discussion}

When analyzing the results from the microgrid scenario study, the statistics show that receiving support from the microgrid for the distribution system has a significant effect on the ${\tt ENS}$. However, the difference between the results of the different scenarios is minor. From the microgrid location study, we know that the placement of the microgrid is an important reliability factor. In addition, since multiple failures can occur at locations where the microgrid is not able to provide support, the distribution system might benefit more if the generation sources were scattered at different locations in the system. 
Another impacting factor is the availability of wind and solar power in the microgrid and the energy level of the battery when a failure occurs. This decides the amount of power the microgrid can contribute.

The reliability of the microgrid, however, is very dependent on the microgrid mode. This is a factor that could change the results if other modes for the microgrid controller are applied. The results indicate that Scenario 3 is the best cross-over strategy for both the microgrid and the distribution system when analyzing ${\tt ENS}$. This could change for the microgrid if other strategies were applied. In addition, the result could be different if other measures are investigated, such as a cost-benefit analysis. 

An interesting result seen from both the distribution system and the microgrid perspective is the increase in SAIFI for the scenarios with support from the microgrid. It was expected that SAIDI would decrease as a result of the microgrid supporting the distribution system since the total duration of load shedding for some load points would decrease. However, based on how SAIFI is measured, the index will increase as a consequence of load points experiencing a new outage period when the microgrid generation units are not able to restore all the supply for the entire outage period. 

The sensitivity analysis revealed some interesting results. The distribution system is very sensitive to the failure rate whereas the microgrid is sensitive to the battery capacity and the repair time. In this scenario, the microgrid is designed to ensure self-sufficiency by storing energy in the battery. For some situations where the repair time is long or multiple failures occur at the same time, the microgrid benefits from a larger battery capacity. 
Several interaction effects are apparent for the microgrid, while none are evident for the distribution system. The two most significant interaction effects seen for the microgrid are linked to the battery capacity, while the battery capacity does not influence the ${\tt ENS}$ for the distribution system. The results from the sensitivity analysis are a consequence of the network sizes. The distribution system is large compared to the microgrid and the load is large compared to the generation in the microgrid. Since the microgrid load is prioritized, the amount of energy support from the microgrid is restricted. In addition, the microgrid can only support faults that are not isolated from the microgrid. Since the distribution system is a relatively large system, local conditions become less severe seen from the whole system perspective. However, they are still important for issues locally as seen from the microgrid perspective. This is also something that could be seen if smaller parts of the distribution system connected to the microgrid were investigated.
The size of the network and distinguishing between local and global perspectives are therefore important factors when analyzing the ${\tt ENS}$ for the given parameters. 
\section{Concluding remarks}
\label{Conclusion}

This study has successfully investigated how a microgrid including RES may improve the reliability of a distribution system. The reliability impact was mapped through investigations of variations in microgrid placement, operation mode, and battery capacity, resulting in a suggestion for the optimal conditions for the microgrid with respect to minimizing the impact on the microgrid while maximizing the contributions to the distribution system. Statistical testing was performed to evaluate the results and indicated a significant difference in receiving support from the microgrid compared to no support. However, the effect could be seen as moderate. The contribution is dependent on the microgrid and the location of the microgrid. Since the microgrid is located in one place, there will be multiple cases where the microgrid is not able to support the distribution network. This is a result of effects seen on the system as a whole against effects seen locally in the system. The significance of the result, however, makes way for further investigations of the possibility of using microgrids or other sources as reliability support for the distribution network. This can lead to further studies of other reliability support possibilities or other system configurations such as distribution networks consisting of multiple microgrids. 
The presented methodology for evaluating how microgrids perform from a reliability perspective using RELSAD shows great promise. The study investigated the reliability of electricity supply both from the perspective of the distribution network and from the perspective of the microgrid. The tool facilitates analysis of multiple different cases and the results are detailed distributions that serve as a good basis for further analysis. Additionally, the tool is general and modular, lowering the barrier to implementing different operation modes and studying different system topologies. 

\section{Acknowledgment}
\label{Acknowledment}

This work is funded by CINELDI - Centre for intelligent electricity distribution, an 8-year Research Centre under the FME-scheme (Centre for Environment-friendly Energy Research, 257626/E20). The authors gratefully acknowledge the financial support from the Research Council of Norway and the CINELDI partners.  
The authors will also like to thank Gerd Kjølle for her helpful contribution and the knowledgeable discussions.

\bibliographystyle{IEEEtran}
\bibliography{References}

\begin{thebibliography}{10}
\providecommand{\url}[1]{#1}
\csname url@samestyle\endcsname
\providecommand{\newblock}{\relax}
\providecommand{\bibinfo}[2]{#2}
\providecommand{\BIBentrySTDinterwordspacing}{\spaceskip=0pt\relax}
\providecommand{\BIBentryALTinterwordstretchfactor}{4}
\providecommand{\BIBentryALTinterwordspacing}{\spaceskip=\fontdimen2\font plus
\BIBentryALTinterwordstretchfactor\fontdimen3\font minus
  \fontdimen4\font\relax}
\providecommand{\BIBforeignlanguage}[2]{{%
\expandafter\ifx\csname l@#1\endcsname\relax
\typeout{** WARNING: IEEEtran.bst: No hyphenation pattern has been}%
\typeout{** loaded for the language `#1'. Using the pattern for}%
\typeout{** the default language instead.}%
\else
\language=\csname l@#1\endcsname
\fi
#2}}
\providecommand{\BIBdecl}{\relax}
\BIBdecl

\bibitem{national2016power}
\BIBentryALTinterwordspacing
{National Academies of Sciences, Engineering, and Medicine and others},
  \emph{The power of change: Innovation for development and deployment of
  increasingly clean electric power technologies}.\hskip 1em plus 0.5em minus
  0.4em\relax National Academies Press, 2016. [Online]. Available:
  \url{https://doi.org/10.17226/21712}
\BIBentrySTDinterwordspacing

\bibitem{IEA2019}
\BIBentryALTinterwordspacing
I.-I.~E. Agency, ``Status of power system transformation 2019.'' [Online].
  Available:
  \url{https://www.iea.org/reports/status-of-power-system-transformation-2019}
\BIBentrySTDinterwordspacing

\bibitem{ipakchi2009}
\BIBentryALTinterwordspacing
A.~Ipakchi and F.~Albuyeh, ``Grid of the future,'' \emph{IEEE Power and Energy
  Magazine}, vol.~7, pp. 52--62, 3 2009. [Online]. Available:
  \url{10.1109/MPE.2008.931384}
\BIBentrySTDinterwordspacing

\bibitem{Prettico2019}
\BIBentryALTinterwordspacing
G.~Prettico and M.~G. Flammini, ``Distribution system operators observatory
  2018 overview of the electricity distribution system in europe,'' 2019.
  [Online]. Available: \url{http://doi.org/10.2760/104777}
\BIBentrySTDinterwordspacing

\bibitem{knowledge2019microgrid}
{Microgrid Knowledge Editorial Team}, ``Microgrid cybersecurity: Protecting and
  building the grid of the future,'' \emph{S\&C Electric Company}, pp. 1--12,
  2019.

\bibitem{Thema}
\BIBentryALTinterwordspacing
{THEMA Consulting Group}, ``Descriptive study of local energy communities,''
  2018. [Online]. Available:
  \url{https://thema.no/wp-content/uploads/THEMA-Reort-2018-20-Local-Energy-Communities-Report-Final.pdf}
\BIBentrySTDinterwordspacing

\bibitem{EUWinter}
\BIBentryALTinterwordspacing
L.~Hancher and B.~Winters, ``The eu winter package briefing paper,'' 2017.
  [Online]. Available:
  \url{https://fsr.eui.eu/wp-content/uploads/The-EU-Winter-Package.pdf}
\BIBentrySTDinterwordspacing

\bibitem{ton2012us}
\BIBentryALTinterwordspacing
D.~T. Ton and M.~A. Smith, ``The us department of energy's microgrid
  initiative,'' \emph{The Electricity Journal}, vol.~25, no.~8, pp. 84--94,
  2012. [Online]. Available: \url{https://doi.org/10.1016/j.tej.2012.09.013}
\BIBentrySTDinterwordspacing

\bibitem{sperstad2020impact}
\BIBentryALTinterwordspacing
I.~B. Sperstad, M.~Z. Degefa, and G.~Kj{\o}lle, ``The impact of flexible
  resources in distribution systems on the security of electricity supply: A
  literature review,'' \emph{Electric Power Systems Research}, vol. 188, p.
  106532, 2020. [Online]. Available:
  \url{https://doi.org/10.1016/j.epsr.2020.106532}
\BIBentrySTDinterwordspacing

\bibitem{mohandes2019review}
\BIBentryALTinterwordspacing
B.~Mohandes, M.~S. El~Moursi, N.~Hatziargyriou, and S.~El~Khatib, ``A review of
  power system flexibility with high penetration of renewables,'' \emph{IEEE
  Transactions on Power Systems}, vol.~34, no.~4, pp. 3140--3155, 2019.
  [Online]. Available: \url{http://doi.org/10.1109/TPWRS.2019.2897727}
\BIBentrySTDinterwordspacing

\bibitem{escalera2018survey}
\BIBentryALTinterwordspacing
A.~Escalera, B.~Hayes, and M.~Prodanovi{\'c}, ``A survey of reliability
  assessment techniques for modern distribution networks,'' \emph{Renewable and
  Sustainable Energy Reviews}, vol.~91, pp. 344--357, 2018. [Online].
  Available: \url{https://doi.org/10.1016/j.rser.2018.02.031}
\BIBentrySTDinterwordspacing

\bibitem{borges2012overview}
\BIBentryALTinterwordspacing
C.~L.~T. Borges, ``An overview of reliability models and methods for
  distribution systems with renewable energy distributed generation,''
  \emph{Renewable and sustainable energy reviews}, vol.~16, no.~6, pp.
  4008--4015, 2012. [Online]. Available:
  \url{https://doi.org/10.1016/j.rser.2012.03.055}
\BIBentrySTDinterwordspacing

\bibitem{celli2013reliability}
G.~Celli, E.~Ghiani, F.~Pilo, and G.~G. Soma, ``Reliability assessment in smart
  distribution networks,'' \emph{Electric Power Systems Research}, vol. 104,
  pp. 164--175, 2013.

\bibitem{mohamad2018development}
\BIBentryALTinterwordspacing
F.~Mohamad, J.~Teh, C.-M. Lai, and L.-R. Chen, ``Development of energy storage
  systems for power network reliability: A review,'' \emph{Energies}, vol.~11,
  no.~9, p. 2278, 2018. [Online]. Available:
  \url{https://doi.org/10.3390/en11092278}
\BIBentrySTDinterwordspacing

\bibitem{kwasinski2012availability}
\BIBentryALTinterwordspacing
A.~Kwasinski, V.~Krishnamurthy, J.~Song, and R.~Sharma, ``Availability
  evaluation of micro-grids for resistant power supply during natural
  disasters,'' \emph{IEEE Transactions on Smart Grid}, vol.~3, no.~4, pp.
  2007--2018, 2012. [Online]. Available:
  \url{http://doi.org/10.1109/TSG.2012.2197832}
\BIBentrySTDinterwordspacing

\bibitem{nosratabadi2017comprehensive}
\BIBentryALTinterwordspacing
S.~M. Nosratabadi, R.-A. Hooshmand, and E.~Gholipour, ``A comprehensive review
  on microgrid and virtual power plant concepts employed for distributed energy
  resources scheduling in power systems,'' \emph{Renewable and Sustainable
  Energy Reviews}, vol.~67, pp. 341--363, 2017. [Online]. Available:
  \url{https://doi.org/10.1016/j.rser.2016.09.025}
\BIBentrySTDinterwordspacing

\bibitem{costa2009assessing}
\BIBentryALTinterwordspacing
P.~M. Costa and M.~A. Matos, ``Assessing the contribution of microgrids to the
  reliability of distribution networks,'' \emph{Electric Power Systems
  Research}, vol.~79, no.~2, pp. 382--389, 2009. [Online]. Available:
  \url{https://doi.org/10.1016/j.epsr.2008.07.009}
\BIBentrySTDinterwordspacing

\bibitem{conti2011generalized}
\BIBentryALTinterwordspacing
S.~Conti, R.~Nicolosi, and S.~Rizzo, ``Generalized systematic approach to
  assess distribution system reliability with renewable distributed generators
  and microgrids,'' \emph{IEEE Transactions on Power Delivery}, vol.~27, no.~1,
  pp. 261--270, 2011. [Online]. Available:
  \url{http://doi.org/10.1109/TPWRD.2011.2172641}
\BIBentrySTDinterwordspacing

\bibitem{conti2014monte}
\BIBentryALTinterwordspacing
S.~Conti and S.~A. Rizzo, ``Monte carlo simulation by using a systematic
  approach to assess distribution system reliability considering intentional
  islanding,'' \emph{IEEE Transactions on Power Delivery}, vol.~30, no.~1, pp.
  64--73, 2014. [Online]. Available:
  \url{http://doi.org/10.1109/TPWRD.2014.2329535}
\BIBentrySTDinterwordspacing

\bibitem{de2017reliability}
\BIBentryALTinterwordspacing
P.~M. de~Quevedo, J.~Contreras, A.~Mazza, G.~Chicco, and R.~Porumb,
  ``Reliability assessment of microgrids with local and mobile generation,
  time-dependent profiles, and intraday reconfiguration,'' \emph{IEEE
  Transactions on Industry Applications}, vol.~54, no.~1, pp. 61--72, 2017.
  [Online]. Available: \url{http://doi.org/10.1109/TIA.2017.2752685}
\BIBentrySTDinterwordspacing

\bibitem{syrri2015contribution}
\BIBentryALTinterwordspacing
A.~Syrri, E.~M. Cesena, and P.~Mancarella, ``Contribution of microgrids to
  distribution network reliability,'' in \emph{2015 IEEE Eindhoven
  PowerTech}.\hskip 1em plus 0.5em minus 0.4em\relax IEEE, 2015, pp. 1--6.
  [Online]. Available: \url{http://doi.org/10.1109/PTC.2015.7232486}
\BIBentrySTDinterwordspacing

\bibitem{farzin2017role}
\BIBentryALTinterwordspacing
H.~Farzin, M.~Fotuhi-Firuzabad, and M.~Moeini-Aghtaie, ``Role of outage
  management strategy in reliability performance of multi-microgrid
  distribution systems,'' \emph{IEEE Transactions on Power Systems}, vol.~33,
  no.~3, pp. 2359--2369, 2017. [Online]. Available:
  \url{http://doi.org/10.1109/TPWRS.2017.2746180}
\BIBentrySTDinterwordspacing

\bibitem{barani2018optimal}
\BIBentryALTinterwordspacing
M.~Barani, J.~Aghaei, M.~A. Akbari, T.~Niknam, H.~Farahmand, and M.~Korp{\aa}s,
  ``Optimal partitioning of smart distribution systems into supply-sufficient
  microgrids,'' \emph{IEEE Transactions on Smart Grid}, vol.~10, no.~3, pp.
  2523--2533, 2018. [Online]. Available:
  \url{http://doi.org/10.1109/TSG.2018.2803215}
\BIBentrySTDinterwordspacing

\bibitem{nikmehr2016reliability}
\BIBentryALTinterwordspacing
N.~Nikmehr and S.~N. Ravadanegh, ``Reliability evaluation of multi-microgrids
  considering optimal operation of small scale energy zones under
  load-generation uncertainties,'' \emph{International Journal of Electrical
  Power \& Energy Systems}, vol.~78, pp. 80--87, 2016. [Online]. Available:
  \url{https://doi.org/10.1016/j.ijepes.2015.11.094}
\BIBentrySTDinterwordspacing

\bibitem{Myhre2022}
\BIBentryALTinterwordspacing
S.~F. Myhre, O.~B. Fosso, P.~E. Heegaard, and O.~Gjerde, ``Relsad: A python
  package for reliability assessment of modern distribution systems,''
  \emph{Journal of Open Source Software}, vol.~7, no.~78, p. 4516, 2022.
  [Online]. Available: \url{https://doi.org/10.21105/joss.04516}
\BIBentrySTDinterwordspacing

\bibitem{RELSAD_documentation}
S.~F. Myhre, ``{RELSAD}-{R}eliability tool for {S}mart and {A}ctive
  {D}istribution systems,'' \url{https://relsad.readthedocs.io/en/latest/},
  2022.

\bibitem{Billinton}
R.~Billinton and R.~N. Allan, \emph{Reliability Evaluation of Power
  Systems}.\hskip 1em plus 0.5em minus 0.4em\relax New York: Springer
  Science+Business Media, LLC, 1996, no. 2nd ed.

\bibitem{943073}
\BIBentryALTinterwordspacing
T.~Langset, F.~Trengereid, K.~Samdal, and J.~Heggset, ``Quality dependent
  revenue caps-a model for quality of supply regulation,'' in \emph{16th
  International Conference and Exhibition on Electricity Distribution, 2001.
  Part 1: Contributions. CIRED. (IEE Conf. Publ No. 482)}, vol.~6, 2001, p. 5
  pp. vol.6. [Online]. Available: \url{https://doi.org/10.1049/cp:20010906}
\BIBentrySTDinterwordspacing

\bibitem{haque1996load}
\BIBentryALTinterwordspacing
M.~Haque, ``Load flow solution of distribution systems with voltage dependent
  load models,'' \emph{Electric Power Systems Research}, vol.~36, no.~3, pp.
  151--156, 1996. [Online]. Available:
  \url{https://doi.org/10.1016/0378-7796(95)01025-4}
\BIBentrySTDinterwordspacing

\bibitem{wang2013new}
\BIBentryALTinterwordspacing
S.~Wang, Z.~Li, L.~Wu, M.~Shahidehpour, and Z.~Li, ``New metrics for assessing
  the reliability and economics of microgrids in distribution system,''
  \emph{IEEE transactions on power systems}, vol.~28, no.~3, pp. 2852--2861,
  2013. [Online]. Available: \url{http://doi.org/10.1109/TPWRS.2013.2249539}
\BIBentrySTDinterwordspacing

\bibitem{t-test}
S.~G. W. and W.~G. Cochran, \emph{Statistical Methods}.\hskip 1em plus 0.5em
  minus 0.4em\relax Iowa State University Press, 1989, no. 8th ed.

\bibitem{KS}
I.~M. Chakravarti, R.~G. Laha, and J.~Roy, \emph{Handbook of methods of Applied
  Statistics Volume I: Techniques of computation, descriptive methods, and
  statistical inference}.\hskip 1em plus 0.5em minus 0.4em\relax John Wiley and
  Sons, 1967.

\bibitem{Stephens1974}
M.~A. Stephens, ``Edf statistics for goodness of fit and some comparisons,''
  \emph{Journal of the American statistical Association}, vol.~69, no. 347, pp.
  730--737, 1974.

\bibitem{Statisticbook}
R.~E. Walpoe, R.~H. Myers, S.~L. Myers, and K.~Ye, \emph{Probability \&
  Statistics for Engineers \& Scientists}.\hskip 1em plus 0.5em minus
  0.4em\relax Pearson Education, Inc, 2011, no. 9th ed.

\bibitem{FASIT}
A.~O. Eggen, ``Fasit kravspesifikasjon,'' Sintef Energy AS., Tech. Rep., 2016,
  (in Norwegian).

\bibitem{baran1989network}
\BIBentryALTinterwordspacing
M.~E. Baran and F.~F. Wu, ``Network reconfiguration in distribution systems for
  loss reduction and load balancing,'' \emph{IEEE Power Engineering Review},
  vol.~9, no.~4, pp. 101--102, 1989. [Online]. Available:
  \url{http://doi.org/10.1109/61.25627}
\BIBentrySTDinterwordspacing

\bibitem{RELSAD}
S.~F. Myhre, ``{RELSAD},'' \url{https://github.com/stinefm/relsad}, 2021.

\bibitem{statnett_2018}
\BIBentryALTinterwordspacing
{Statnett SF}, ``Yearly statistics 2018 {(in Norwegian)},'' 2018. [Online].
  Available:
  \url{https://www.statnett.no/for-aktorer-i-kraftbransjen/systemansvaret/leveringskvalitet/statistikk/}
\BIBentrySTDinterwordspacing

\end{thebibliography}


\end{document}